\documentclass[conference]{IEEEtran}
\IEEEoverridecommandlockouts
\usepackage{cite}
\usepackage{amsmath,amssymb,amsfonts}
\usepackage{algorithm}
\usepackage{graphicx}
\usepackage{textcomp}
\usepackage{multirow}
\usepackage[hyphens,spaces,obeyspaces]{url} 
\usepackage{hyperref}
\usepackage{amsmath}
\usepackage{float}
\usepackage[noend]{algpseudocode} 
\usepackage{booktabs}
\usepackage{easyReview}
\usepackage{tabularx}
\usepackage{listings}
\usepackage{xcolor}

\usepackage{caption}
\usepackage{subcaption}
\usepackage[export]{adjustbox}[2011/08/13]

\usepackage{etoolbox}
\makeatletter
\patchcmd{\@maketitle}
  {\addvspace{0.5\baselineskip}\egroup}
  {\addvspace{-2\baselineskip}\egroup}
  {}
  {}
\makeatother

\makeatletter
\newcommand{\linebreakand}{%
  \end{@IEEEauthorhalign}
  \hfill\mbox{}\par
  \mbox{}\hfill\begin{@IEEEauthorhalign}
}
\makeatother

\newcolumntype{L}{>{\centering\arraybackslash}m{2cm}}
\newcolumntype{S}{>{\centering\arraybackslash}m{1cm}}
\newcolumntype{M}{>{\centering\arraybackslash}m{1.5cm}}

\definecolor{codegreen}{rgb}{0,0.6,0}
\definecolor{codegray}{rgb}{0.5,0.5,0.5}
\definecolor{codepurple}{rgb}{0.58,0,0.82}
\definecolor{backcolour}{rgb}{1,1,1}
\lstdefinestyle{mystyle}{
    backgroundcolor=\color{backcolour},   
    commentstyle=\color{codegreen},
    keywordstyle=\color{magenta},
    numberstyle=\tiny\color{codegray},
    stringstyle=\color{codepurple},
    basicstyle=\ttfamily\footnotesize,
    breakatwhitespace=false,         
    breaklines=true,                 
    captionpos=t,                    
    keepspaces=true,                 
    numbers=left,                    
    numbersep=5pt,                  
    showspaces=false,                
    showstringspaces=false,
    showtabs=false,                  
    tabsize=2
}
\algdef{SE}[VARIABLES]{Variables}{EndVariables}
   {\algorithmicvariables}
   {\algorithmicend\ \algorithmicvariables}
\algnewcommand{\algorithmicvariables}{\textbf{global variables}}

\def\BibTeX{{\rm B\kern-.05em{\sc i\kern-.025em b}\kern-.08em
    T\kern-.1667em\lower.7ex\hbox{E}\kern-.125emX}}
\begin{document}

\title{A Machine Learning Approach Towards Runtime Optimisation of Matrix Multiplication\\
\vspace{-0.5em}
}

\author{\IEEEauthorblockN{Yufan Xia}
\IEEEauthorblockA{\textit{Australian National University}\\
Canberra, Australia \\
xiayufan12345@outlook.com}
\and
\IEEEauthorblockN{Marco De La Pierre}
\IEEEauthorblockA{\textit{Pawsey Supercomputing Research Centre} \\
Perth, Australia \\
marco.delapierre@pawsey.org.au}
\linebreakand

\IEEEauthorblockN{Amanda S. Barnard}
\IEEEauthorblockA{\textit{Australian National University}\\
Canberra, Australia \\
amanda.s.barnard@anu.edu.au}
\and 
\IEEEauthorblockN{Giuseppe Maria Junior Barca}
\IEEEauthorblockA{\textit{Australian National University}\\
Canberra, Australia \\
giuseppe.barca@anu.edu.au}\\
}

\maketitle

\begin{abstract}

The GEneral Matrix Multiplication (GEMM) is one of the essential algorithms in scientific computing. Single-thread GEMM implementations are well-optimised with techniques like blocking and autotuning. However, due to the complexity of modern multi-core shared memory systems, it is challenging to determine the number of threads that minimises the multi-thread GEMM runtime.

We present a proof-of-concept approach to building an Architecture and Data-Structure Aware Linear Algebra (ADSALA) software library that uses machine learning to optimise the runtime performance of BLAS routines. More specifically, our method uses a machine learning model on-the-fly to automatically select the optimal number of threads for a given GEMM task based on the collected training data. Test results on two different HPC node architectures, one based on a two-socket Intel Cascade Lake and the other on a two-socket AMD$^{\circledR}$ Zen 3, revealed a 25 to 40 per cent speedup compared to traditional GEMM implementations in BLAS when using GEMM of memory usage within 100 MB.

\end{abstract}


\section{Introduction}

In the past three decades, significant effort has been devoted to improving the General Matrix Multiplication (GEMM) implementation \cite{drevet2011optimization,dalton2015optimizing,valero2020slass,yang2021libshalom,allen_note_2009,nath_improved_2010,peise_performance_2012,low_analytical_2016,
Drevet2011}. As one of the essential linear algebra subroutines in the Basic Linear Algebra Subprograms (BLAS) collection, the optimisation of GEMM is crucial to many HPC applications \cite{blackford2002updated}. Previous work on optimising GEMM yielded considerable speedup over the historical approach before the 1990s when software engineers had to manually tune their BLAS operations. {The automatically tuned linear algebra software (ATLAS) and its improvements are the first batch of auto-tuning efforts on optimising linear algebra operations; they are able to auto-tune the linear algebra operation codes by searching over parameters like blocking factor, loop order, and partial storage location on each specific computer architecture \cite{whaley1998automatically,gunnels2001family,vuduc2001statistical}. Later in the 2000s, the self-optimised linear algebra routine (SOLAR) attempted the analytical modelling of the timing of multi-process linear algebra operations \cite{gunnels2001systematic,valsalam2002framework,demmel2005self,herrero2006framework,cuenca2002towards,Cuenca2004}. }


{There is little work on using machine learning (ML) or similar approaches to optimise single or multi-threaded linear algebra routines. In 2001, Vuduc \emph{et al.} proposed a Support Vector Classifier to model single-thread linear algebra performance and choose the best tiling strategy among caches and memories \cite{vuduc2001statistical}. The polynomial regression has been attempted to optimise the number of threads or block size used based on empirical features and a speed comparable to the Intel$^{\circledR}$ MKL Cholesky decomposition was obtained with PLASMA\footnote{Parallel Linear Algebra for Scalable Multi-Core Architectures} \cite{camara2013empirical}. Previous research by Peise \emph{et al.} adopted a polynomial regression to model the dense linear algebra run times, and they applied additional techniques to boost the performance of this simple model \cite{peise2012performance}. }

Most of the prior work in this area has focused on improving GEMM with large and regular-shaped input matrices, while not as much success has been attained in addressing the performance of GEMM with small and irregular-shaped input matrices, even though they are frequently used in many applications of scientific computing software and machine learning methods \cite{yang2021libshalom,frison2018blasfeo}.  For example, ResNet's convolution kernel uses GEMM with matrix operands of size $ 64\times 3000$ \cite{he2016deep,yang2021libshalom}.




In addition, the widespread adoption of multi-core CPUs  has made the optimisation of multi-thread GEMM implementations crucial. Matrix blocking improves GEMM performance by increasing the cache hit rate during the GEMM execution; for multi-thread GEMM implementations, blocking is also used for thread-wise job assignments \cite{catalan2019case}. Although for single-threaded GEMM, extensive work has been carried out to optimize performance by tuning the size of matrix blocks on different system architectures,  less frequent have been the efforts in optimising multi-thread GEMM through parameter tuning \cite{Katagiri2017}.

In this article, we focus on the runtime optimisation of existing multi-thread BLAS GEMM implementations, such as those in the Intel$^{\circledR}$ MKL and AMD$^{\circledR}$ BLIS libraries, by providing a strategy to automatically select the number of threads that minimises execution time. This differs from, and is synergistic with previous work, which mainly focused on optimising the GEMM kernels.  

\begin{figure}
  \centering
  {\includegraphics[width=180pt]{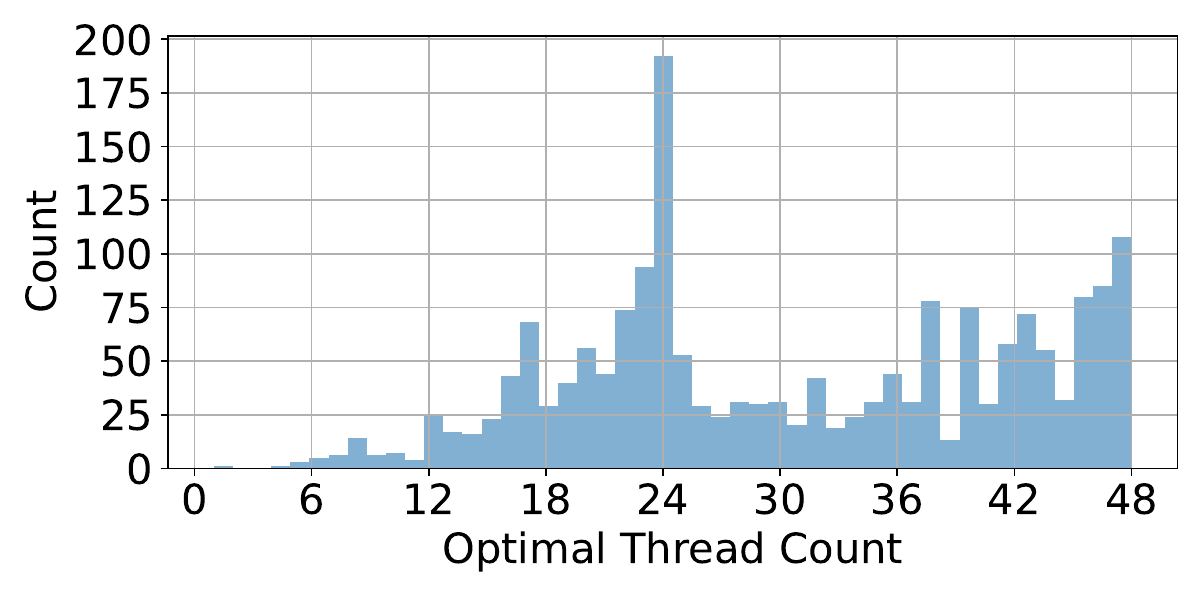}}  
    \vspace{-1em}  
  \caption{Histogram of the optimal thread count; measured with single precision GEMM with memory usage within 100 MB. The experiment was run on a single HPC node with 2 Cascade Lake CPUs and 48 physical cores. the BLAS package used is Math Kernel Library.}
  \label{fig:core_hist} 
  \vspace{-2em}  
\end{figure}

Figure \ref{fig:core_hist} shows a histogram of the optimal thread numbers when running single-precision GEMM using the MKL package for matrices with an aggregate memory footprint lower than 100 MB on a compute node of the Gadi supercomputer with a two-socket Intel$^{\circledR}$ Cascade Lake architecture with a total of 48 physical cores (detailed in subsection \ref{subsec: Experimentation Platforms}). GEMM users commonly use the maximum available number of threads (48 in this case) to minimise GEMM runtime. However, as shown in Fig. \ref{fig:core_hist}, thread counts lower than 48 often provide better GEMM wall-time, with potential wall-time savings up to 70 per cent of average, as later shown in subsection \ref{subsec: Data Collection}. This experiment revealed a large room for improvements by suitable selection of the number of threads at runtime.



 However, choosing the number of threads that minimises the GEMM execution time is challenging due to the underlying diversity and complexity of modern shared-memory computer architectures \cite{chandra2001parallel,chennupati2017scalable}. {Recently, to address this issue Luan \emph{et al.} \cite{Luan2022} proposed a thread auto-tuning system at runtime for iterative parallel programs, although this has not been specifically tested on GEMM.}
 
 Here we propose an ML-based approach to tackle this problem. For a given GEMM input configuration and computer system architecture, this approach uses an ML model to select the number of threads that minimises the GEMM runtime. The software conducts benchmark experiments upon installation to obtain data for training the ML model. The training of the ML model itself is performed at installation time and is specifically tailored to the computer system architecture in use. Then, the ML model is evaluated at runtime to predict the optimal number of threads to use. Since our software uses GEMM implementations as black boxes, it is synergistic with any existing optimisations. 
 Our software implementation enables us to speed up GEMM, notwithstanding the runtime overhead of ML evaluations. On the Gadi supercomputer located at the National Computational Infrastructure (NCI) and on the Setonix supercomputer located at the Pawsey Supercomputing Centre (please refer to section \ref{sec: Experimentation Information} for details about NCI and Pawsey), our approach achieved an average of $26\%$ and $41\%$ reduction in runtime. 


The remaining of this article is organised as follows. Section \ref{sec: Background} introduces the background for GEMM and ML; in Section \ref{sec: Software Workflow}, the software design is proposed, then in Section \ref{sec: Machine Learning Methods} the ML methods are detailed. Section \ref{sec: Experimentation Information} briefs the experimentation platform and settings; Section \ref{sec: Performance Analysis} shows and analyses the performance of the ML models and software speedup. Section \ref{sec: Conclusion} concludes, and future research is outlined.


\section{Background}
\label{sec: Background}
In this Section, we introduce the GEMM routine, ML algorithms, and data preprocessing techniques applied throughout.



\subsection{The GEMM routine}
The GEMM routine is a third-level BLAS operation. A standard GEMM interface is shown in Listing \ref{sgemm interface} \cite{blackford2002updated}.
 Its parameters include two input matrices (arrays) $A$ and $B$ of $m$ by $k$ and $k$ by $n$ sizes, one output matrix $C$ of $m$ by $n$ size, the dimensions of the matrices $m$, $k$, and $n$, and additional parameters such as the multiplicative factors $\alpha$ and $\beta$, flags for transpose, and previous values of matrix $C$.  

Given matrices $A$, $B$, and $C$, GEMM performs the following operation
\begin{equation}\label{equ: GEMM}
    {C} \leftarrow \alpha {A} {B}+\beta {C}
\end{equation}

For multi-thread GEMM implementations, the number of threads can be set using environment variables or function calls outside the standard GEMM interface. In our runtime optimisation, we consider the matrix dimensions' effect and the number of threads used on the GEMM running time.

\begin{minipage}{\columnwidth}
\begin{center}
    \vspace{1em} 
    \begin{lstlisting}[frame=tb,language={C}, caption={Interface of SGEMM (single precision).}, label={sgemm interface}, numbers=none,captionpos=b,linewidth=8cm]

sgemm (int   TRANSA, int TRANSB,
       int        M, int      N, int      K,
       float  ALPHA, float    A, int    LDA,
       float      B, int    LDB, float BETA,
       float      C, int    LDC)
    \end{lstlisting}
    \vspace{0em}
\end{center}
\end{minipage}

\subsection{Machine Learning Algorithms}

In order to develop a high-performance ML model, a selection must be carried out across suitable candidates. We briefly describe the dataset we will be working on (data gathering is detailed in subsection \ref{subsec: Data Gathering}) to motivate the choice of the candidate models with the most suitable characteristics. The dataset typically contains $10^3$ data points with 10-20 dimensions; the distribution of most data features is skewed. It is a relatively small dataset with low data dimensions, but the mapping between features and the label might be complex since we expect this to be a non-linear relation due to the polynomial time complexity of GEMM operations and the complexity of multi-core computer architectures. Knowing these characteristics, we introduce several ML algorithms to select candidates, as shown in Table \ref{tab:model_comp}.  
\begin{table}[t]
  \centering
  \caption{Comparisons of ML model characteristics.}
  \label{tab:model_comp}
  
{
\scriptsize
\sffamily

\begin{tabular}{SLSMM}

\toprule
\\[-2ex]
{\textbf{Model Catagories}} & {\textbf{Models}} &  {\textbf{Parametric}} & {\textbf{Good with Data Imbalance}} & {\textbf{Data Size Requirement}} \\
\midrule
\multirow{ 5}{0.15\columnwidth}{\textbf{Linear Models}} & Linear Regression & \multirow{ 5}{0.15\columnwidth}{Yes} & \multirow{ 5}{0.15\columnwidth}{No} & \multirow{ 1}{0.15\columnwidth}{Medium} \\
\cmidrule(r){2-2}\cmidrule(r){5-5}
& ElasticNet &  &  & \multirow{ 1}{0.15\columnwidth}{Medium} \\
\cmidrule(r){2-2}\cmidrule(r){5-5}
& Bayesian Regression &  &  & \multirow{ 1}{0.15\columnwidth}{Small} \\
\midrule
\multirow{ 8}{0.15\columnwidth}{\textbf{Tree \\Based Models}} & Decision Tree & \multirow{ 8}{0.15\columnwidth}{No} & \multirow{ 8}{0.15\columnwidth}{Yes} & \multirow{ 8}{0.15\columnwidth}{Medium}\\
\cmidrule(r){2-2}
& XGBoost &  &  &  \\
\cmidrule(r){2-2}
& AdaBoost &  &  &  \\
\cmidrule(r){2-2}
& Random Forest &  &  &  \\
\cmidrule(r){2-2}
& LightGBM &  &  &  \\
\midrule
\multirow{ 3}{0.15\columnwidth}{\textbf{Other Models}} & SVM Regressor & \multirow{ 3}{0.15\columnwidth}{No} & \multirow{ 3}{0.15\columnwidth}{No} & \multirow{ 1}{0.15\columnwidth}{Small}\\
\cmidrule(r){2-2}\cmidrule(r){5-5}
& KNN Regressor &  & &\multirow{ 1}{0.15\columnwidth}{Medium}\\

\bottomrule
\end{tabular}
}

    \vspace{-1em}
\end{table}

Linear algorithms produce simple ML models with common advantages, such as their high speed in training and evaluation \cite{deisenroth2020mathematics}. However, they suffer from low predictive accuracy for non-linear mappings. Due to the nature of our task, evaluation speed is an essential factor, and linear ML models may enable fast real-world GEMM speedup. Linear algorithms have different regularisation capabilities to prevent over-fitting, so we included linear regression, ElasticNet and Bayesian regression as our candidates \cite{tibshirani1996regression,bishop2003bayesian}. 

Additionally, we include both Decision Tree and tree ensemble models as candidates. The Decision Tree algorithm is non-parametric, it uses a set of rules to map each data instance to a discrete class or a continuous value \cite{bishop2006pattern}. The Random Forest is an ML algorithm that trains multiple Decision Trees in parallel using subsets of the training set. It evaluates by averaging the result of its child predictors \cite{breiman2001random,geron2019hands}. AdaBoost, XGBoost and LightGBM are ML algorithms that train child predictors serially; each new child predictor is designed to correct its predecessor's error \cite{chen2016xgboost,freund1996experiments}. These tree ensemble algorithms generally can reduce bias and variances compared to a single Decision Tree and thus can produce a more promising result but require more time to evaluate.
\begin{figure*}[!htbp]
    \centering
    {\includegraphics[width=13cm]{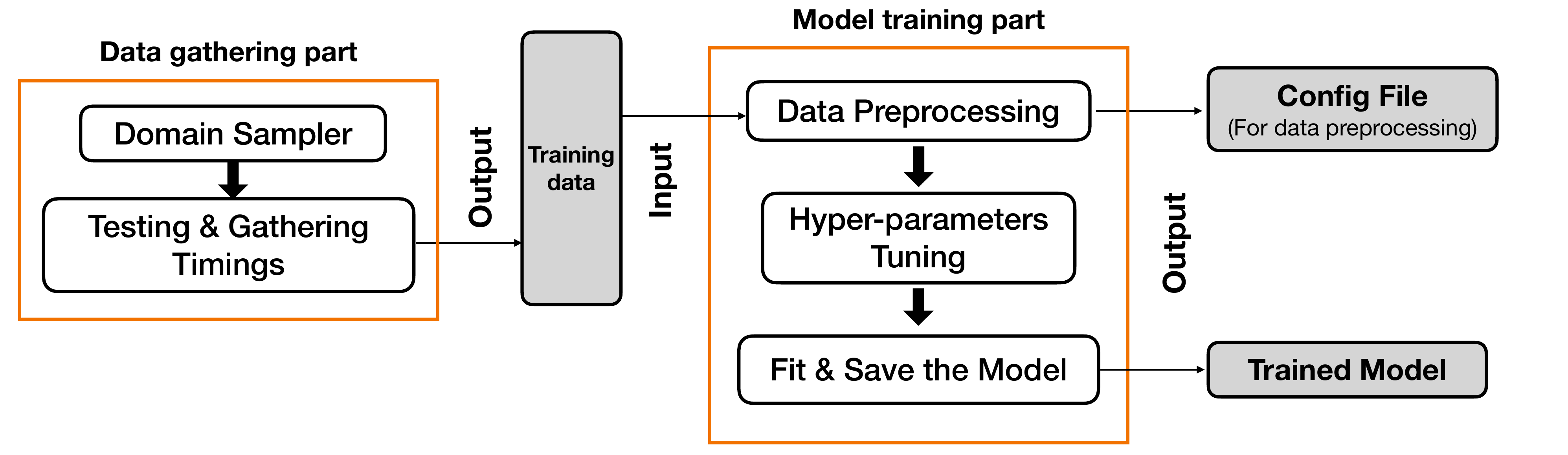}}
    \caption{The installation workflow of ADSALA GEMM. Upon ADSALA installation, the library performs two sub-parts shown in the diagram. In the end, two files containing the configurations together with the production-ready ML model will be saved for later use at runtime.}
    \label{fig:Lib Structure}
  \end{figure*}
\begin{figure}[!htbp]
  \centering
    \vspace{-1em}
  \includegraphics[width=0.7\columnwidth]{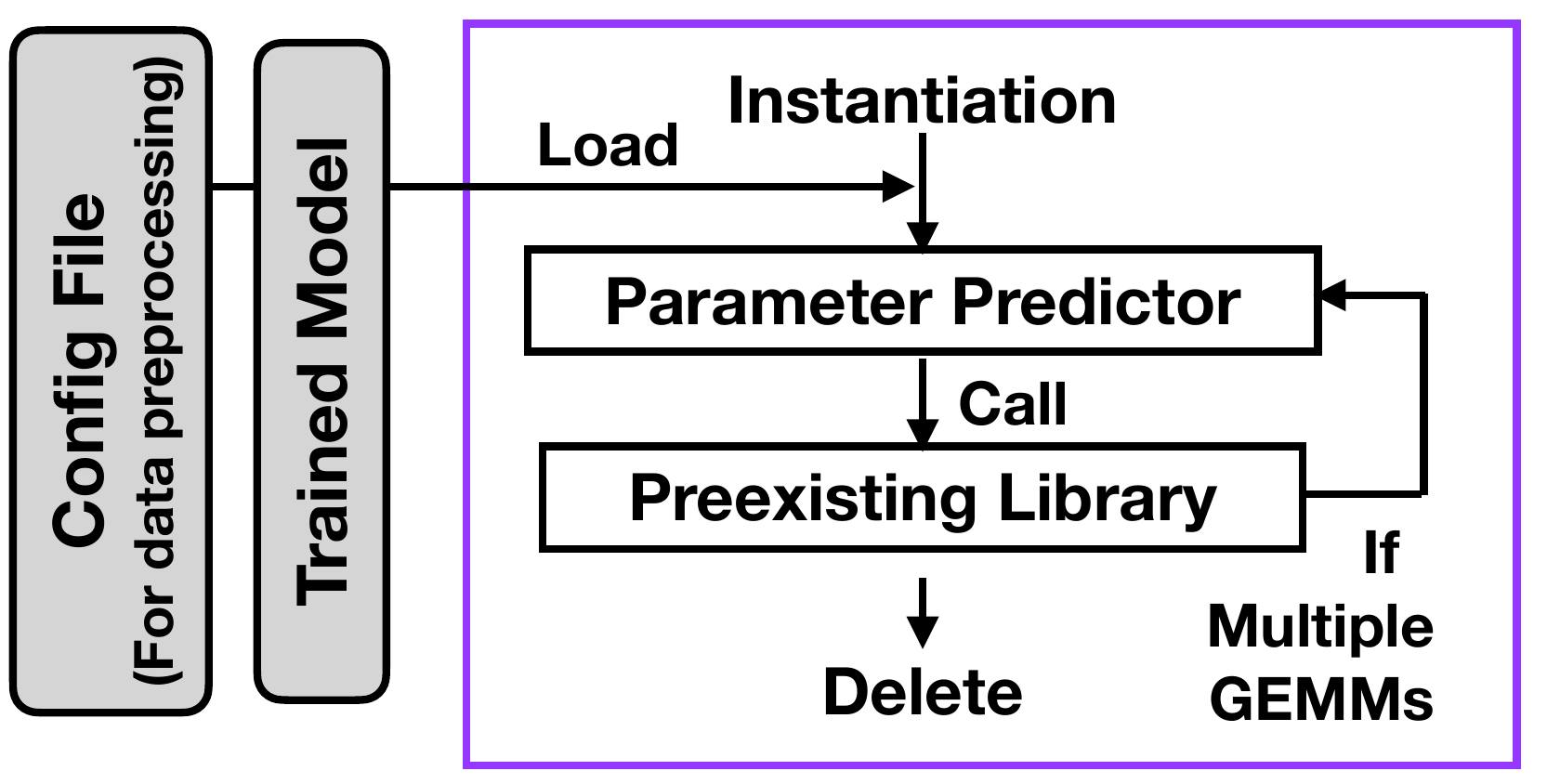} 
  \caption{The runtime workflow of ADSALA GEMM. Configuration file and trained ML model output during installation (see Fig. \ref{fig:Lib Structure}) are used by the runtime library. } 
  \label{fig: Predictor Design} 
    \vspace{-1em}
\end{figure}

Support Vector Machine (SVN) and k-Nearest-Neighbors (kNN) are non-parametric algorithms but are designed to work well for high dimensions and unknown data structures, respectively \cite{awad2015support}. Since we have low data dimensionality, they might not demonstrate their advantages when learning from our dataset; kNN is also known to be slow in evaluation \cite{bhatia2010survey}. Thus they are unsuitable for our use case.

\subsection{Data Preprocessing Techniques}
\label{subsec: Data Preprocessing Techniques}
Removing local outliers can help improve the ML model's predictive performance.  We apply the Local Outlier Factor (LOF) method to the dataset. Statistical methods for outlier removal are sufficient for removing global outliers that do not fit in the global data point distribution but can not recognise local outliers that do not fit in their local neighbourhood data distribution \cite{han2022data}. LOF provides a method for identifying local outliers \cite{10.1145/335191.335388}. LOF is a density-based method; it assigns each data point a degree of isolation from its surrounding points, and from this, the local outliers can be filtered.

The Yeo-Johnson transformation is a data transformation technique used in regression that can improve the predictive performance of the ML model by remapping the feature distribution to near-Gaussian. In contrast, the original Box-Cox feature transformation requires all feature values to be positive before transformation, which is not always satisfactory \cite{sakia1992box}. Some modified versions of Box-Cox accept non-positive values, but the automatic estimation of parameter values is often unstable \cite{sakia1992box}. Yeo-Johnson improves from the Box-Cox transformation and preserves the advantages of Box-Cox while including non-positive feature values and also provide a stable parameter estimation \cite{weisberg2001yeo}. The parameter $\lambda$ in the Yeo-Johnson transformation controlling the transformation effect can be automatically derived for each feature from the original data distribution through maximum likelihood estimation (MLE). We apply Yeo-Johnson with MLE for automating the ML workflow.


\section{Software Workflow}
\label{sec: Software Workflow}

\subsection{Overarching Workflow}
On each platform, runtime data for GEMM can be gathered that is associated with each specific combination of the number of threads, input dimensions, and BLAS package. This data can be used to train an ML model that can accurately predict the number of threads that gives us the shortest runtime for given $m$, $k$, and $n$ dimensions of the matrices in the GEMM input. The final product is an Architecture and Data Structure Aware Linear Algebra (ADSALA) library that selects the optimal number of threads at runtime, adapting to different HPC platforms with different system architectures and BLAS packages. 

Figure \ref{fig:Lib Structure} and \ref{fig: Predictor Design} illustrate the procedures of our software design. The software is divided into a data gathering part, a model training part and a runtime library. The first two parts commence at the installation time of the software, while the third part is dedicated to runtime usage when the user program links to our library. These three parts are designed to produce a suitable ML model fitted to the data collected on the target machine and a high-speed runtime ML model evaluation to improve GEMM execution time.

\subsection{Installation Workflow}
Upon installation, the software gathers the training data by experimentation. During the data gathering part, the software first quasi-randomly samples from the domain of dimensions of GEMM inputs. Then these sampled domains are passed to a timing program for running the corresponding GEMM operations and timing collection. The collected timings are stored in a file to serve as the training data for the ML model. 

The gathered data is then preprocessed to be ready for ML model fitting. Afterwards, the hyper-parameters of the ML model to be used at runtime are tuned using the preprocessed data.

To account for the statistical noise of GEMM runtime, each GEMM input is executed multiple times during the timing data collection step. Unexpected changes in runtime can be observed when changing the number of threads at runtime, which decreases the accuracy of the GEMM runtime collection. For minimum interference in data gathering, we avoid changing the number of threads at runtime by separating experiments with different numbers of threads to different program execution. Speed-wise, the data collection can be easily applied on multiple exclusive compute nodes to shorten the wall-time for installation. 

\subsection{Runtime Workflow}
\label{subsec: Runtime Workflow}
 The ML model and its configuration are loaded into  memory at the program boot time. During the program runtime, the user program calls GEMM as a function. The ML model then evaluates the optimal number of threads on-the-fly and runs the GEMM implementation using that specific thread number.
 
 The ML model and configuration are wrapped in a C++ class; before the class instance is destroyed, the GEMM can be executed multiple times without reloading the required files. After the last call of GEMM finishes, the class instance holding the ML model can be safely destroyed to free the memory space.
 
In many applications, GEMM usage is within a loop with the same GEMM input size, and multiple ML model evaluations of the same GEMM matrix dimensions are redundant. The software is designed to remember the last GEMM input and ML predictions; if the current GEMM matrix dimensions are the same as the previous, the software will read and apply the predictions from the responsible class attributes without re-evaluation. 


\section{Machine Learning Methods}
\label{sec: Machine Learning Methods}

\subsection{Mechanism for Predictions}
For a given combination of $m$, $k$, and $n$ values, the software selects the optimal number of threads by first predicting the runtime associated with the possible number of threads. {From the list of predicted runtimes, the number of threads that provides the shortest runtime is chosen for the ensuing GEMM calculation.} In this way, the regression ML model outputs the runtime of GEMM rather than the number of threads we need to use with GEMM. 


\subsection{Data Gathering}
\label{subsec: Data Gathering}
The memory usage of GEMM is $4 (mk+kn+mn)$ Bytes for single precision (SGEMM) and $8 (mk+kn+mn)$ Bytes for double precision (DGEMM); the upper bound to GEMM dimensions ($m$, $k$, and $n$) is limited by the available memory size (we do not consider matrix sizes bigger than memory). 

Since we use GEMM with matrices of all shapes and sizes within the memory limits, including slim/square and big/small matrices, the sampled domains need to be evenly distributed across the space. Thus, we use a scrambled Halton sequence to generate a low discrepancy quasi-random sequence for the data sampling \cite{MascagniChi+2004+435+442}. Since the samples have multiple dimensions, we use the scrambled Halton sequence rather than the regular Halton sequence to mitigate the correlation between dimensions \cite{MascagniChi+2004+435+442}. We use bases 2, 3, and 4 as the basis of the sequence generation for dimensions $m$, $k$, and $n$, respectively. These samples are sent to the timing program as GEMM input parameters to collect their runtime. 


\subsection{Feature Engineering and Data Preprocessing}
\label{subsec: Data Preprocessing}

Table \ref{tab:additional features} shows the features used for the ML model. The features are divided into two groups; Group 1 is associated with serial runtime terms and Group 2 with parallel runtime terms. 

{The GEMM runtime is an architecture-dependent function of combinations of the matrix dimensions $m$, $n$ and $k$, and the number of threads, $n\_threads$. In Group 1, $m*k$, $k*n$, $m*n$, and $m*k+k*n+m*n$ represent the size of matrices $A$, $B$, $C$, and the total memory size in single-precision words, respectively, which are known to have a direct relation with the number of memory operations and therefore with the GEMM runtime. Specifically, these terms tend to dominate the serial runtime for small matrix dimensions. The cubic term $m*k*n$ is proportional to the number of floating-point operations performed that tend to dominate runtime in serial execution for large matrix dimensions. In parallel, the FLOP workloads are parallelized over threads, leading to terms like ${m*n*k}/n\_threads$.}

{In the feature selection phase, we generated many candidate features as combinations of $m$, $n$, $k$ and $n\_threads$, and then removed those with high correlation among each other, leading to the final set of selected features shown in Table \ref{tab:additional features}.}

\begin{table}[t!]
\vspace{-1.5em}
  \centering
  \caption{List of available features.}
  \label{tab:additional features} 
  
{
    \scriptsize
    \sffamily
    
    \begin{tabular}{p{0.05\textwidth}cc}
    \toprule
    &\textbf{Group 1} & \textbf{Group 2} \\
    \midrule
    1 & m &    ${\text{m}}/{\text{n\_threads}}$\\[3pt]
    2 & k&    ${\text{k}}/{\text{n\_threads}}$\\[3pt]
    3 & n&    ${\text{n}}/{\text{n\_threads}}$\\[3pt]
    4 & n\_threads&    ${\text{m*k}}/{\text{n\_threads}}$\\[3pt]
    5 & m*k&    ${\text{m*n}}/{\text{n\_threads}}$\\[3pt]
    6 & m*n&    ${\text{k*n}}/{\text{n\_threads}}$\\[3pt]
    7 & k*n&    ${\text{m*k*n}}/{\text{n\_threads}}$\\[3pt]
    8 & m*k*n&    ${(\text{m*k+k*n+m*n})}/{\text{n\_threads}}$\\[3pt]
    9 & m*k+k*n+m*n& \\
    \bottomrule    
    \end{tabular}
}


  \vspace{-1em}
\end{table}

Since the training data size is sufficient for model validation, we use cross validation folds rather than the leave-one-out method in the hyper-parameter tuning process to reduce its computational cost \cite{geron2019hands}. Stratified sampling is used in the train-test set split and validation set split to ensure a similar distribution in the train set, test set, and validation sets \cite{geron2019hands}. 
 
\begin{figure}[ht!]
  \centering 
  {\includegraphics[width=\columnwidth]{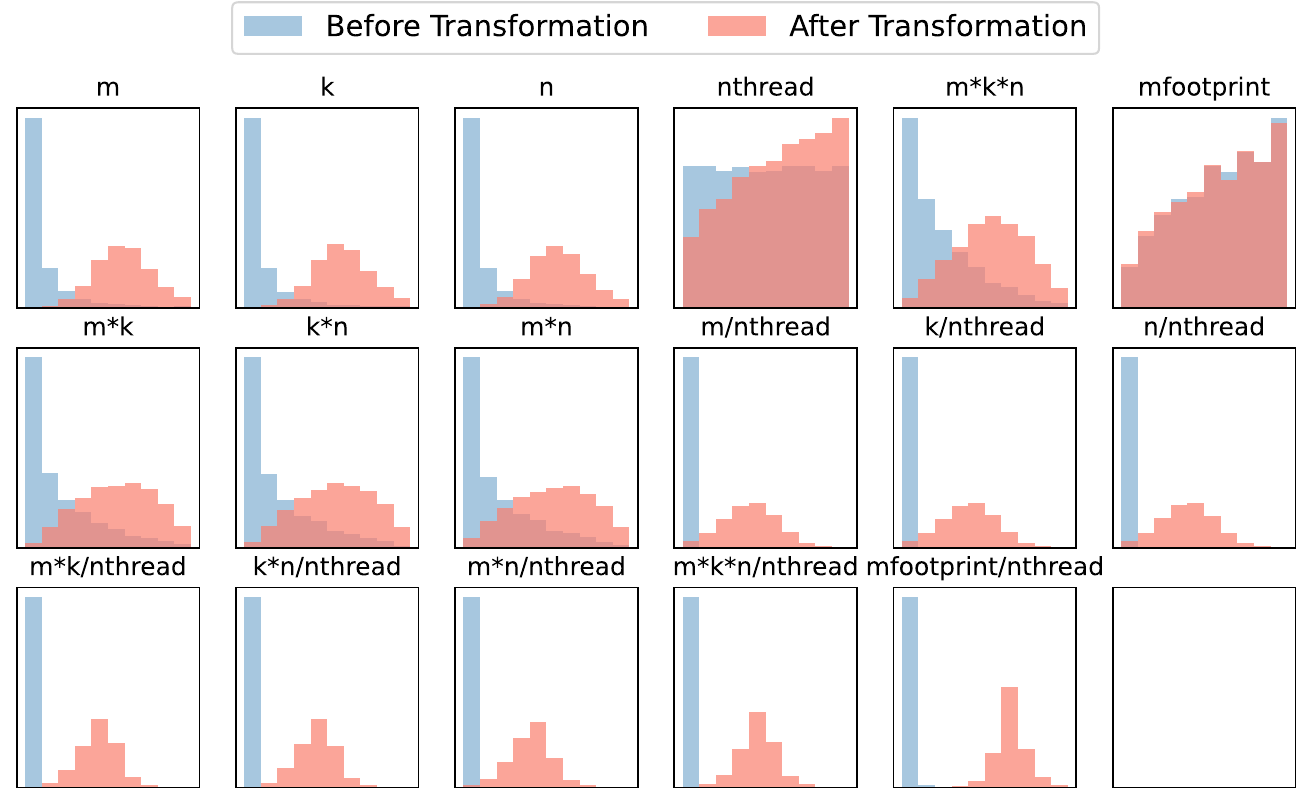}}
  \vspace{-1.5em}
  \caption{{Feature distributions before and after transformation. The data is from GEMM with input matrix size below 500 MB on Setonix.}}
\vspace{-1em}
  \label{fig:before_feature_distribution}
\end{figure}

Our GEMM input space produces skewed distributions in most features in the sampled data; an example is presented in Fig. \ref{fig:before_feature_distribution}. This imbalanced distribution will cause machine learning methods to struggle to describe the relationships between the features and the label \cite{geron2019hands}. We apply {the} Yeo-Johnson transformation (introduced in subsection \ref{subsec: Data Preprocessing Techniques}) to map their distribution to near-Gaussian distributions; this transformation estimates the most suitable $\lambda$ (as introduced in \ref{subsec: Data Preprocessing Techniques}) values and controls the transformation effect on different features through MLE method \cite{weisberg2001yeo}. After the transformation, we perform a standardisation process on features to ensure all features are on a similar scale \cite{geron2019hands}.  

Furthermore, we use LOF (introduced in subsection \ref{subsec: Data Preprocessing Techniques}) to remove global and local outliers \cite{10.1145/335191.335388}. The outlier removal with LOF is applied after the standardisation because LOF is a density-based method and thus requires a similar scale in all dimensions. We then remove features with correlation coefficients with other features larger than a threshold of $80\%$. For each correlation feature pair, we remove the feature with the larger total correlation with the other features. Afterwards, the hyper-parameter tuning is performed for all models before the model selection.

\subsection{Model Selection}
\label{subsec: Model Tuning and Selection}

There are two main criteria for selecting the ML model that maximises the speed improvement: the predictive performance and the evaluation speed (the model evaluation time must be negligible compared to the GEMM runtime to obtain better speedup). As discussed earlier, depending on the model evaluation overhead, the model with the best predictive accuracy may not yield the optimal runtime speedup. Vice versa, if a model features fast evaluation runtime but low predictive accuracy, the resulting GEMM speedup would be limited as well. 




Our strategy for selecting the final model that combines these two criteria is as follows. We model the speedup by: $$s = \frac{t_{\text{original}}}{t_{\text{ADSALA}}+t_{\text{eval}}}$$ where $t_{\text{ADSALA}}$ is the GEMM runtime using predicted number of threads, $t_{\text{eval}}$ is the ML model evaluation time on the target machine, and $t_{\text{original}}$ is the GEMM runtime using the maximum number of available threads. We can measure $t_{\text{eval}}$ of each tuned model by averaging multiple runs on the target HPC platform. By averaging the estimated speedup $s$ across all GEMM in our test dataset, the ML model yielding the highest average speedup is selected as the most suitable model.


\section{Experimentation Information}
\label{sec: Experimentation Information}
This Section introduces the supercomputing platforms used for experimentation and the experimentation setup on those platforms.

\subsection{Experimentation Platforms}
\label{subsec: Experimentation Platforms}
Our experiments were conducted on two supercomputing platforms.

\subsubsection{Setonix} 

\begin{figure}[h!]
  \vspace{-1em}
  \centering
  {\includegraphics[width=0.7\columnwidth]{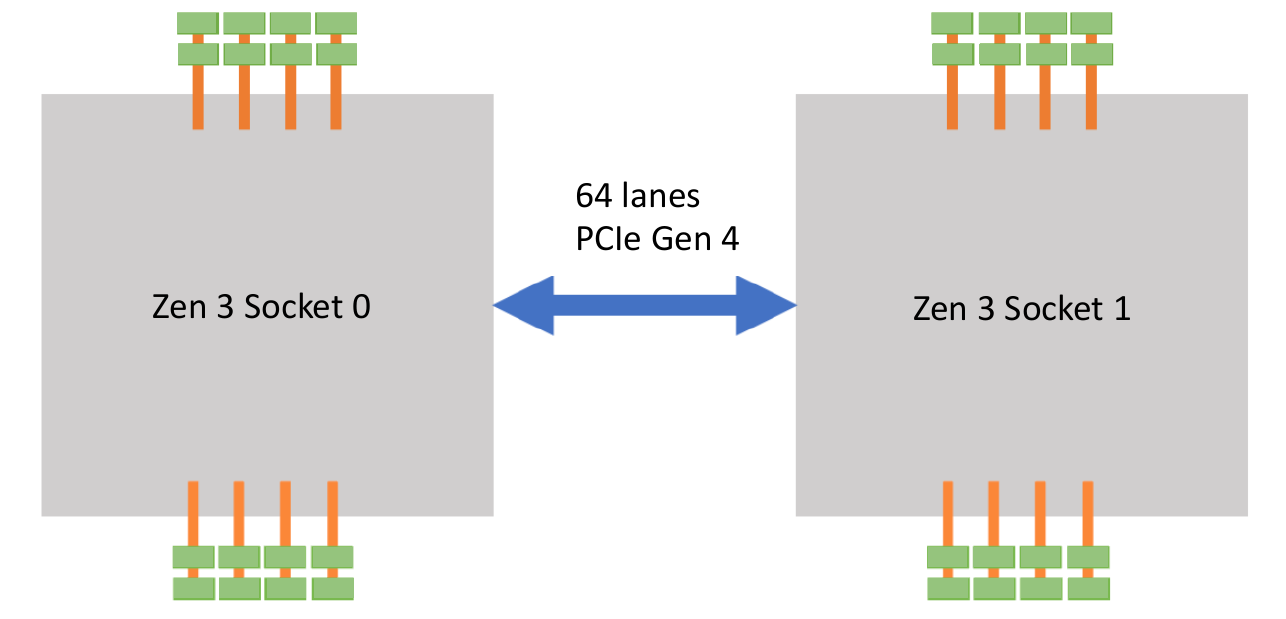}}
  \caption{A schematic diagram for the 2-socket EPYC CPU configuration on Setonix.}
  \label{fig: Setonix Socket}
  \vspace{-1em}  

\end{figure}

Setonix is a supercomputer located at the Pawsey Supercomputing Research Centre in Australia. As shown in Fig. \ref{fig: Setonix Socket}, each of its compute nodes has specifics:
\begin{itemize}
  \item Two CPU sockets with AMD$^{\circledR}$ EPYC 64-Core \textit{Milan} CPU (2.55 GHz). Hyper-threading is supported, allowing 256 simultaneous threads per compute node. Each \textit{Milan} CPU has eight modules, and each of these modules contains eight Zen 3 cores and an exclusive 32 MB level three cache shared among these eight cores.
  \item 256GB of Memory as eight NUMA domains, with four NUMA domains per socket; eight memory channels per socket are supported.
\end{itemize}

\subsubsection{Gadi}

\begin{figure}[h!]

  \centering
  {\includegraphics[width=0.9\columnwidth]{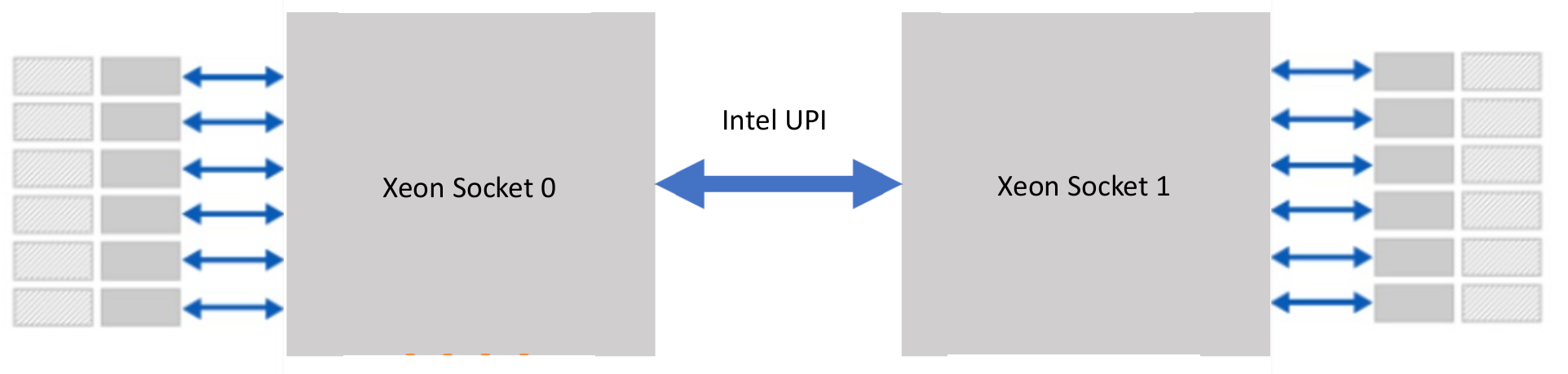}}
  \caption{A schematic diagram for 2-socket Cascade Lake CPU configuration; sockets are connected using Intel$^{\circledR}$ UPI (Ultra Path Interconnect). }
  \label{fig: Gadi Socket}
  \vspace{-1em}
\end{figure}

Gadi is a supercomputer located at the Australian National Computational Infrastructure. As shown in Fig. \ref{fig: Gadi Socket}, Each of Gadi's compute nodes has the following specifics:
\begin{itemize}
  \item Two CPU sockets with Intel$^{\circledR}$ Xeon 24-Core \textit{Cascade Lake} CPU (Platinum 8274, 3.2 GHz). Hyper-threading is supported, allowing 96 simultaneous threads per compute node.
  \item 192GB of Memory as four NUMA domains, with two NUMA domains per socket; six memory channels per socket are supported.
  \end{itemize}

\subsection{Experimentation Setup}
AMD$^{\circledR}$ recommends using BLIS\footnote{ \url{https://developer.amd.com/amd-aocl/blas-library/ }} on their processors; thus, on Setonix, we will use the performance of BLIS as the baseline. The CPU of Gadi is Intel$^{\circledR}$-based, and Intel$^{\circledR}$ recommends using MKL\footnote{ \url{https://www.intel.com/content/www/us/en/develop/documentation/get-started-with-mkl-for-dpcpp/top.html} } on their processors, so we will use the performance of MKL package as the baseline for measuring performance improvements. 

\subsubsection{Node Access}
The experimentation is done on one compute node as we are interested in multi-thread GEMM that runs on shared-memory architectures. We set exclusive node access (even for jobs with less than the maximum number of threads) to eliminate the effect on GEMM timing from other compute jobs. 

\subsubsection{NUMA Memory Policy} 
 To comply with Intel$^{\circledR}$'s benchmark settings\footnote{\url{https://www.intel.com/content/www/us/en/developer/articles/technical/benchmarking-gemm-with-intel-mkl-and-blis-on-intel-processors.html}}, on Setonix and Gadi, we set the NUMA policy for memory to \textit{interleave} for all threads, enforcing a round robin algorithm for the memory allocation where a new NUMA domain is used after the previous NUMA domain is fully used. Our testing found that setting the NUMA memory policy stabilises the GEMM runtime.
  
\subsubsection{Timing measurements and byte alignment}
To measure timings reliably, ten iterations of the same-size GEMM are performed in a loop structure, as done in the data collection. These GEMM input matrices are allocated using \textit{memalign} in and filled with random numbers; the alignment of the matrices is set to 64 bytes to assist optimal usage of vector instructions.
   
\subsubsection{Thread Affinity} 
\begin{figure}[htbp]
  \vspace{-1em}
  \centering
  {\includegraphics[width=0.85\columnwidth]{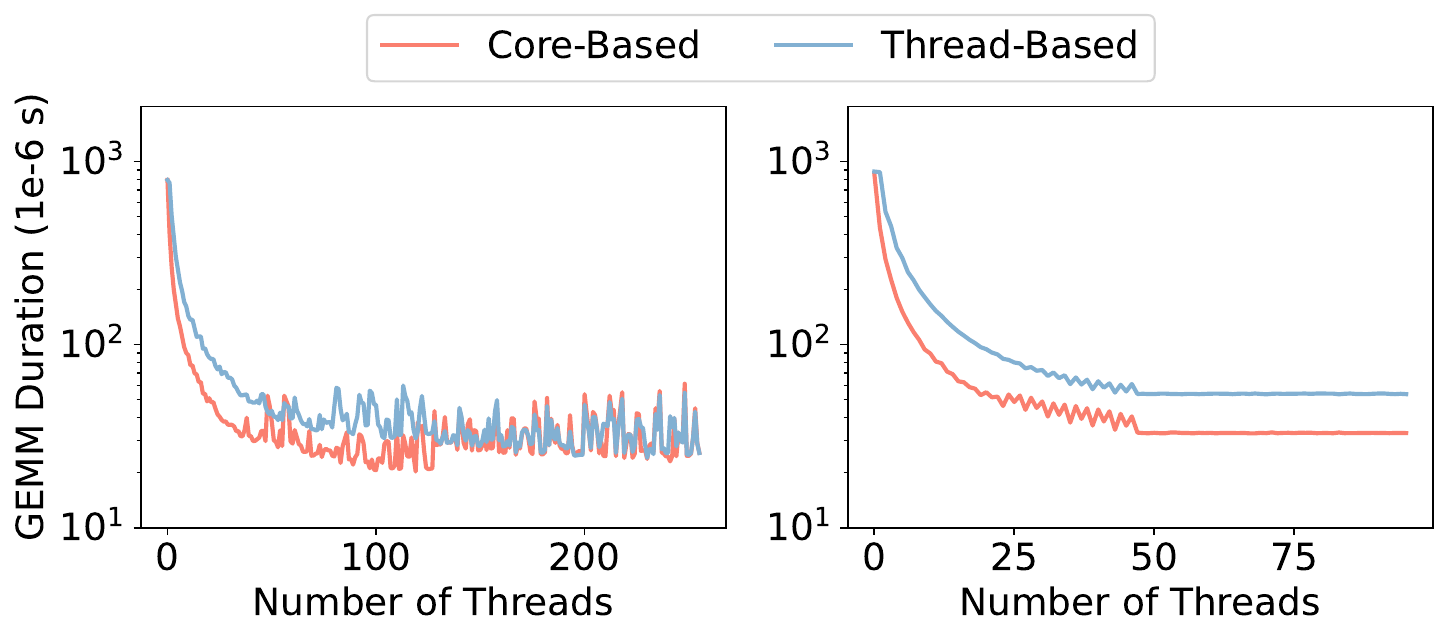}} 
    \vspace{-0.7em}
  \caption{Thread Affinity Comparison on Setonix (left) and Gadi (right); tested GEMM is quasi-randomly generated with memory usage within 500 MB; the axis for GEMM duration is plotted in log scale to illustrate the difference. } 
  \label{fig: thread_affinity} 
\end{figure}

 OpenMP is used for setting the thread affinity policy for a better and more stable GEMM performance. Figure \ref{fig: thread_affinity} shows the average runtime over the test dataset when using core-based  ($OMP\_PLACES=cores$) and thread-based ($OMP\_PLACES=threads$)
  affinity settings, respectively.
We can observe that the core-based thread affinity setting provides a shorter GEMM execution time when the number of threads is less than half of the maximum number for both Gadi and Setonix. On Setonix, core-based affinity yields better performance for up to 128 threads. At the same time, there is only a minor performance difference between the two affinity settings when the number of threads is larger than 128. On Gadi, the performance advantage of core-based affinity  is observed across the whole range of the number of threads. Thus, we used a core-based affinity setting for all our computational experiments. 

\section{Performance Analysis}
\label{sec: Performance Analysis}
In this Section, we visualise and briefly explain our datasets, present the ML model selection result, and then analyze the performance of our method with its software implementation on a test GEMM dataset with random dimensions and a test GEMM dataset with predesigned dimensions.

The results discussed in this Section focus solely on SGEMM (single precision GEMM). Thus, for conciseness in this Section we use interchangeably the SGEMM and GEMM terms.

\subsection{Datasets{, Training Time}, and Data Visualization}
\label{subsec: Data Collection}

The datasets gathered on Setonix and Gadi both contain 1763 different GEMM inputs with a memory upper bound of 500 MB. {Learning curves for the training and validation loss were built to determine how much data was necessary to train an accurate machine learning model. From the learning curves, it was observed that 1763 GEMM samples are sufficient for matrices below 500 MB, as more training data did not lead to a significant increase in the validation performance.}

{The data gathering required 112 node hours on Setonix (single node), while constructing the ML model with XGBoost took 4 node hours. The data gathering on Gadi took around 6 hours on fifteen nodes and XGBoost model construction took around 4 hours on one compute node (the resource consumption is not measured in node hours on Gadi).} 

\begin{figure}[!htbp]   
  \centering
    \vspace{-0.5em}
  {\includegraphics[width=0.7\columnwidth]{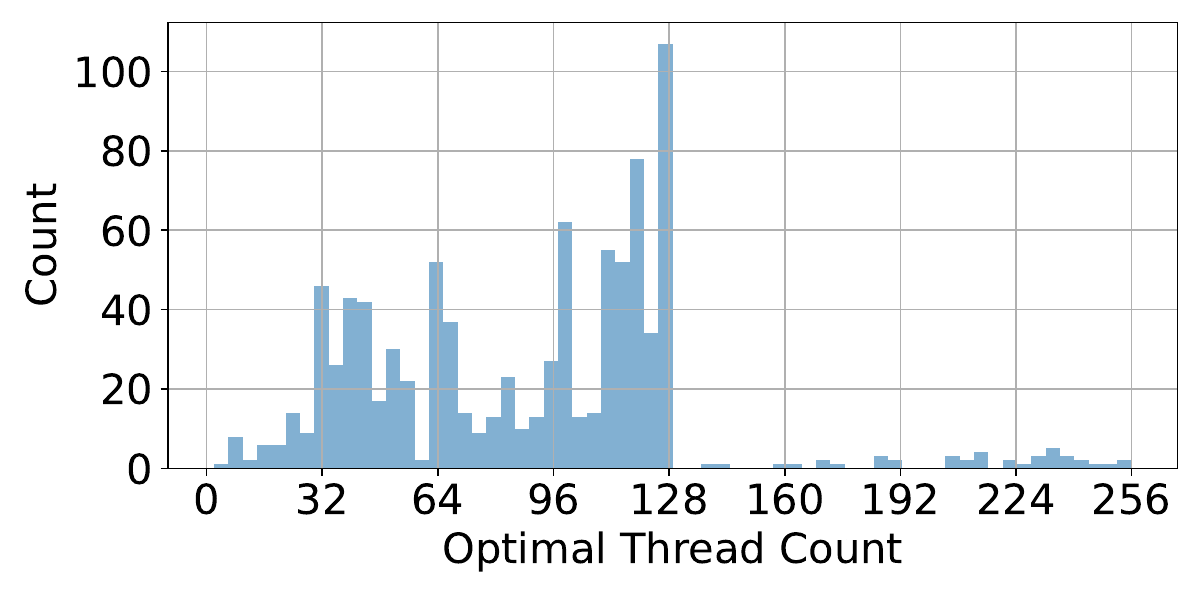}}
  \caption{Histogram of the optimal number of threads with at least one of $m,k,n$ smaller than $1{,}000$; the data is from Setonix with memory usages within 500 MB.} 
  \label{fig: Setonix_1000_dis} 
\end{figure} 

\begin{figure}[!t]
  \centering 
  \vspace{-1em}
     \begin{subfigure}[b]{1\columnwidth}
         \centering
        \makebox[\columnwidth][c]{\includegraphics[width=1.06\columnwidth]{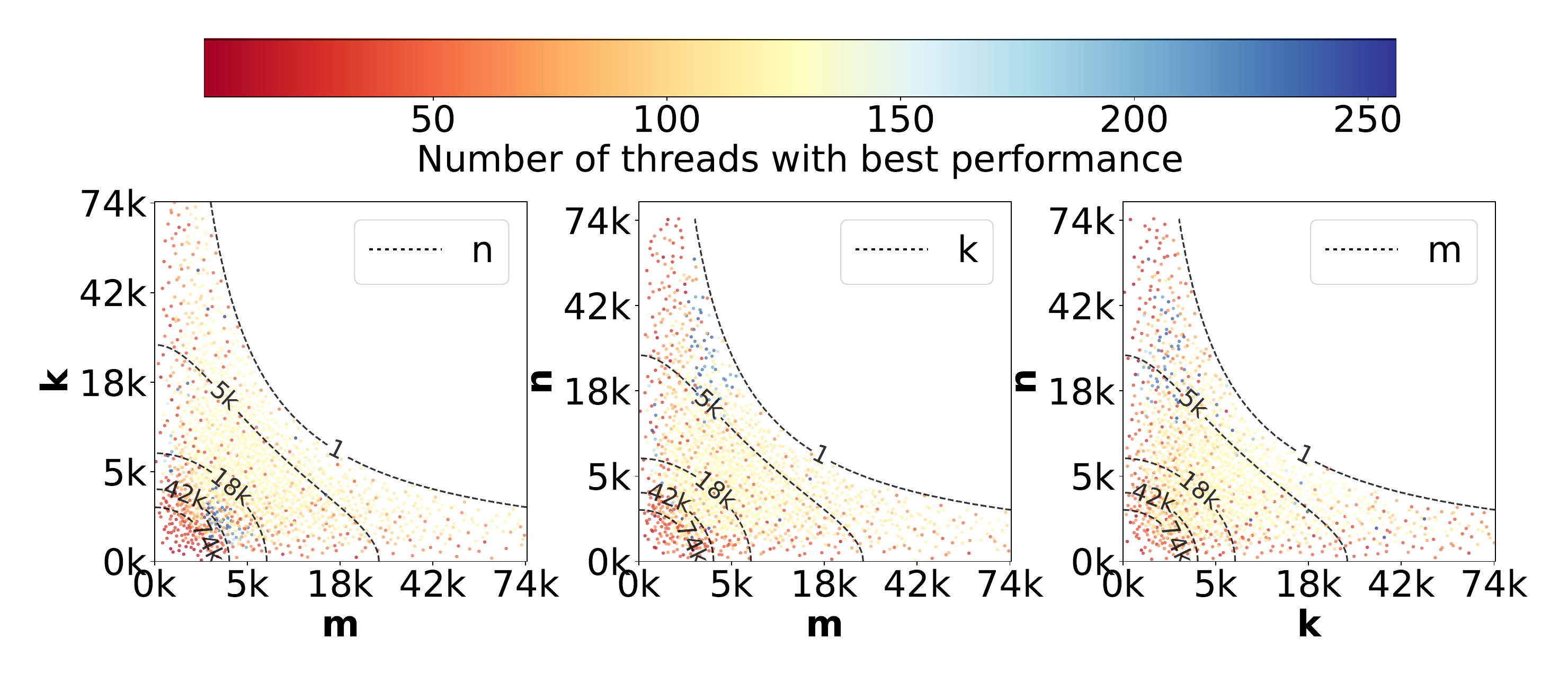}}
        \vspace{-2.2em}
        \subcaption{Setonix}
        \label{fig: data collection Setonix} 
     \end{subfigure}
    \vspace{-1.5em}
     \begin{subfigure}[b]{1\columnwidth}
         \centering
        \makebox[\columnwidth][c]{\includegraphics[width=1.06 \columnwidth]{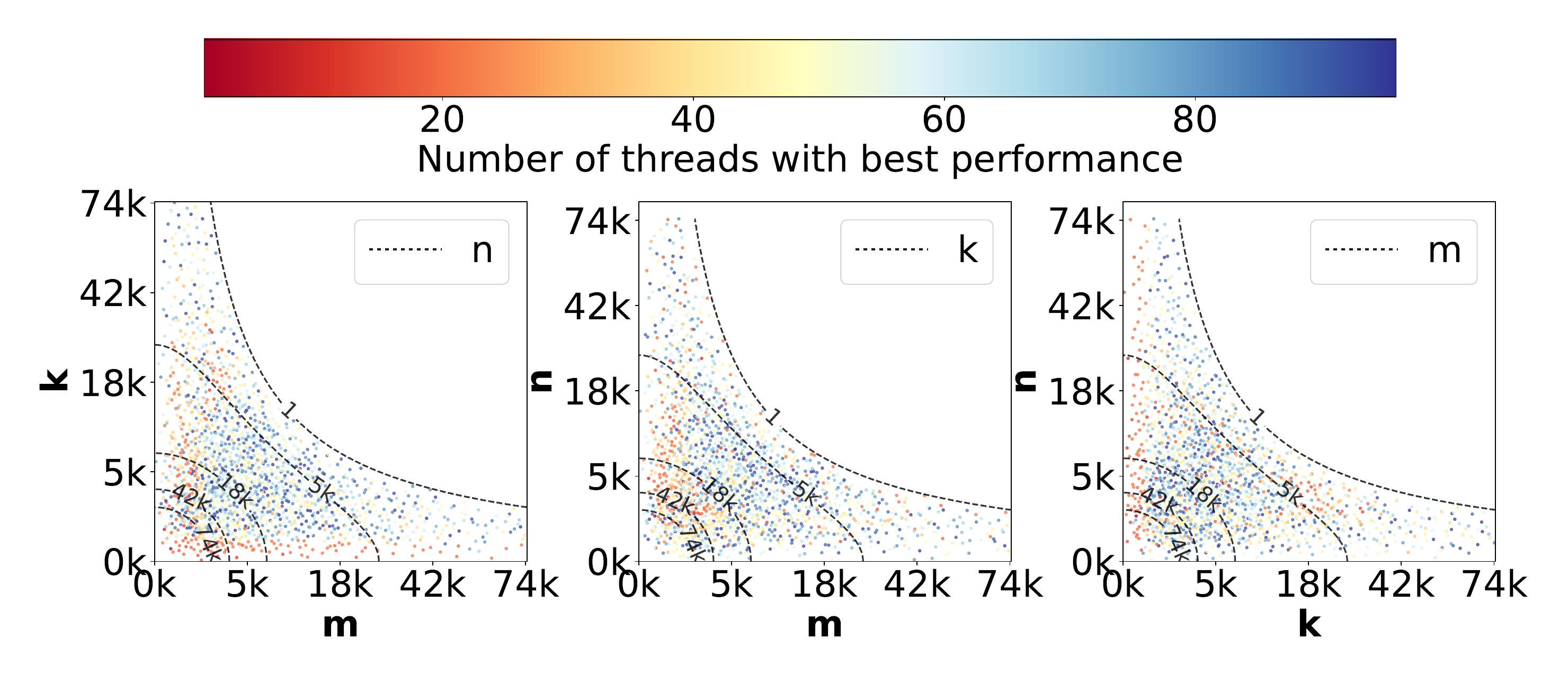}}
        \vspace{-2.2em}
        \subcaption{Gadi}
        \label{fig: data collection Gadi} 
        \vspace{1.5em}
     \end{subfigure}

  \caption{Heatmap of the optimal number of threads on Setonix and Gadi. The horizontal and vertical axes use a square root scale. The dashed lines on each sub-graph are contour lines of the sampling domain with each label showing the value of the third dimension.} 
  \label{fig: data collection Gadi} 
   \vspace{-1.8em}
\end{figure}

The distribution of the optimal number of threads on Setonix with BLIS GEMM is shown in Fig. \ref{fig: data collection Setonix}. The data from Setonix demonstrates that using as many threads as the number of logical cores available frequently leads to sub-optimal GEMM performance. For GEMMs with at least one dimension among $m$, $k$, and $n$ smaller than $1{,}000$, the fastest number of threads tends to be less than half of the maximum available number of threads, as shown in Fig. \ref{fig: Setonix_1000_dis}. 
\begin{table*}[!htbp]
  \vspace{-1em}
  \caption{Model performance and estimated speedups for ML models on Setonix.}
   \centering
  \label{tab:estimated_speedup_setonix}
  \scriptsize
\sffamily

\begin{tabular}{crLLLLLL}
    \toprule
    {} &  \textbf{Model} &    \textbf{Normalised Test RMSE} &  \textbf{Ideal mean speedup} &  \textbf{Ideal aggregate speedup}  &  \textbf{Model evaluation time in $\mu$s} &  \textbf{Estimated mean speedup} &  \textbf{Estimated aggregate speedup} \\
    \midrule
     &  \textbf{Linear Regression} &       0.97 &                1.19 &                   1.03  &           42.94 &                    1.18 &                       1.03 \\
     &  \textbf{ElasticNet} &     1.00 &                1.08 &                   0.90  &           24.98 &                    1.08 &                       0.90 \\
     &  \textbf{Bayes Regression} &       0.97 &                1.19 &                   1.03  &            7.91 &                    1.19 &                       1.03 \\
     &  \textbf{Decision Tree} &       0.28 &                0.50 &                   0.50  &            11.07 &                   0.50 &                   0.50 \\
     &  \textbf{Random Forest} &       0.18 &                1.58 &                   1.43  &        20816.25 &                    0.76 &                       0.83 \\
     &  \textbf{AdaBoost} &       0.42 &                0.55 &                   0.57  &          133.11 &                    0.55 &                       0.56 \\
     &  \textbf{XGBoost} &       {0.13} &                {1.50} &                   1.38  &          45.02 &                    1.50 &                       1.37 \\
     &  \textbf{LightGBM} &   0.40 & 1.31  & 1.21   &  57.60   &   1.30 &  1.21\\
    \bottomrule
\end{tabular}
\end{table*}

\begin{table*}[!htbp]
  \caption{Model performance and estimated speedups for ML models on Gadi.}
   \centering 
  \vspace{-0.5em}
  \label{tab:estimated_speedup_gadi}
  \scriptsize
\sffamily

\begin{tabular}{crLLLLLL}
    \toprule
    {} &  \textbf{Model} &    \textbf{Normalised Test RMSE} &  \textbf{Ideal mean speedup } &  \textbf{Ideal aggregate speedup}  &  \textbf{Model evaluation time in $\mu$s} &  \textbf{Estimated mean speedup} &  \textbf{Estimated aggregate speedup} \\
    \midrule
     &  \textbf{Linear Regression} &       0.96 &                0.98 &                   0.97 &          106.88 &                    0.98 &                       0.96 \\
     &  \textbf{ElasticNet} &     1.00 &                1.00 &                   0.99 &           53.79 &                    0.99 &                       0.99 \\
     &  \textbf{Bayes Regression} &       0.96 &                0.98 &                   0.97 &           18.89 &                    0.98 &                       0.97 \\
     &  \textbf{Decision Tree} &       0.18 &                0.88 &                   0.87 &            8.99 &                    0.88 &                       0.87 \\
     &  \textbf{Random Forest} &       0.08 &                1.05 &                   1.03 &         3240.34 &                    0.90 &                       0.94 \\
     &  \textbf{AdaBoost} &       0.29 &                0.67 &                   0.70 &          129.34 &                    0.67 &                       0.70 \\
     &  \textbf{XGBoost} &       0.05 &                1.07 &                   1.04 &           54.11 &                    1.06 &                       1.03 \\
     &  \textbf{LightGBM} &   0.12 &               1.03 &                   1.01 &           70.70 &                    1.03 &                       1.01 \\
    \bottomrule
\end{tabular}
  \vspace{-1em}
\end{table*}

For GEMM inputs with larger sizes and squarer matrix shapes, the optimal number of threads tends to be roughly half the maximum. Qualitatively speaking, the optimal number of threads distributes in a mostly symmetric pattern for GEMM inputs $m$, $k$, and $n$, except for the small cluster of data points showing a novel pattern by having their optimal number of threads close to the maximum (shown as the blue cluster in Fig. \ref{fig: data collection Setonix}).

The optimal number of threads when using MKL GEMM on Gadi is shown in Fig. \ref{fig: data collection Gadi}.  Compared to Setonix, a different pattern of the best-performing number of threads with respect to $m$, $k$, and $n$ can be observed; for example, there is a higher portion of GEMM having their optimal number of threads near the maximum number of threads. The asymmetric pattern in the optimal number of threads in terms of $m$, $k$, and $n$ is also different. However, there are also similarities with the results obtained on Setonix; for example, the GEMM inputs with larger sizes and squarer matrix shapes have their optimal number of threads closer to the maximum. 

These results illustrate the complex performance patterns yielded by multi-thread GEMM; our ML models are expected to learn these patterns and choose the best number of threads. 

As discussed in Section \ref{subsec: Data Preprocessing}, the data set splitting for model training and testing is performed using stratified sampling, with the fraction of the test set to 30 $\%$. 

\subsection{Model Performance and Selection}

 Our candidate models were trained and tuned using the collected and preprocessed data, and tested on the test set as discussed in the previous Section.

 Tables \ref{tab:estimated_speedup_setonix} and \ref{tab:estimated_speedup_gadi} show a series of metrics including the test set Root Mean Squared Error (RMSE), the estimated speedup ignoring model evaluation time, the model evaluation time, and the estimated speedup when including model evaluation time for Setonix and Gadi, respectively. The aggregate speedup is calculated as the total wall-time of ADSALA GEMM for the test set divided by the total wall-time of traditional GEMM using a number of threads equal to the number of cores for the test set. Similarly, the mean speedup is calculated by averaging the speedup of each ADSALA GEMM over a traditional GEMM contained in the test set.

The data in Tables \ref{tab:estimated_speedup_setonix} and \ref{tab:estimated_speedup_gadi} clearly show that the ML model evaluation overhead and its accuracy could both severely impinge on the attainable speedup. In general, although linear ML models have several to hundreds of times shorter evaluation times than tree-based algorithms, but they also offer lower predictive performance than the latter. Random Forest, LightGBM, and kNN have relatively low RMSE, their slow evaluation speed causes a drastic decrease in the estimated speedup, making them unable to provide any improvement of the GEMM runtime (speedup below one). 

As shown in Tables \ref{tab:estimated_speedup_setonix} and \ref{tab:estimated_speedup_gadi}, ML models with better predictive performance (lower RMSE) often yield a higher estimated speedup, with some exceptions such as linear regression compared to ElasticNet on Gadi, and AdaBoost on both platforms. 

On both platforms, XGBoost has the lowest evaluation time among tree-based algorithms and the best predictive performance, resulting in the best estimated speedup. Therefore, the XGB Regressor was selected as the ML method of choice for both platforms.




 

\subsection{Performance Assessment using Software Implementation}

\begin{table}[!htbp]
  \centering
  \vspace{-0.5em}
  \caption{Speedup statistics on Setonix and Gadi with hyper-threading}
  \label{tab: performance}
  \scriptsize
\sffamily

\begin{tabular}{LSSSS}
    \toprule
     {} & \textbf{Setonix 0-500 MB} & \textbf{Setonix 0-100 MB} & \textbf{Gadi 0-500 MB} & \textbf{Gadi 0-100 MB}\\
    \midrule
    {\textbf{${\text{Mean Speedup}}$}}              &    1.32  &     {1.41}     &   1.07    &   1.26             \\
    {\textbf{${\text{Standard Deviation}}$}}& {0.41}    &   0.21     &     0.70     &    0.28                \\
    \midrule
    {\textbf{${\text{Min Speedup}}$}}               &  0.76    &   {0.87}       &   0.88    &    0.97                \\
    {\textbf{${\text{25th Percentile}}$}}   &   1.05  &   {1.17}        &   1.00    &      1.01           \\
    {\textbf{${\text{50th Percentile}}$}}   &  1.18   &    {1.29}       &   1.00    &      1.17              \\
    {\textbf{${\text{75th Percentile}}$}}   &  1.37     &   {1.61}      &   1.02    &       1.40             \\
    {\textbf{${\text{Max Speedup}}$}}      &    9.05      &    {2.49}        &   3.01    &       1.98                 \\
    \bottomrule
\end{tabular} 
  \vspace{-1em}
\end{table}

\begin{table}[!htbp]
  \centering
  \vspace{-0.5em}
  \caption{{Speedup statistics on Setonix and Gadi with no hyper-threading}}
  \label{tab: performance HT off}
  \scriptsize
\sffamily

\begin{tabular}{LSSSS}
    \toprule
     {} & \textbf{Setonix 0-500 MB} & \textbf{Setonix 0-100 MB} & \textbf{Gadi 0-500 MB} & \textbf{Gadi 0-100 MB}\\
    \midrule
    {\textbf{${\text{Mean Speedup}}$}}              &    1.24  &     {1.55}     &   1.02    &   1.34             \\
    {\textbf{${\text{Standard Deviation}}$}}& {0.58}    &   0.72     &     0.24     &    0.83                \\
    \midrule
    {\textbf{${\text{Min Speedup}}$}}               &  0.23    &   {0.60}       &   0.52    &    0.55                \\
    {\textbf{${\text{25th Percentile}}$}}   &   0.91  &   {1.16}        &   0.96    &      0.95           \\
    {\textbf{${\text{50th Percentile}}$}}   &  1.09   &    {1.39}       &   0.99    &      1.08              \\
    {\textbf{${\text{75th Percentile}}$}}   &  1.41     &   {1.75}      &   1.01    &       1.42             \\
    {\textbf{${\text{Max Speedup}}$}}      &    6.15      &    {8.54}        &   3.88    &       7.53                 \\
    \bottomrule
\end{tabular} 
  \vspace{-1em}
\end{table}

The results discussed in this Section use an additional data set independent from the training and test data sets used for the ML model training and evaluation. 

This additional data set includes 174 data points sampled using a scrambled Halton sequence within the GEMM input domains with memory requirements within 500 MB. This ensures that the data used for the performance analysis of our ADSALA GEMM software is sampled using a more uniform, low discrepancy pattern within the testing domain.

Using the runtime with the number of threads set equal to the number of cores as a reference, we tested the speedup of our ADSALA software implementation over the 174 GEMM within the new low-discrepancy data set.

The statistics of the ADSALA GEMM speedup results are shown in Table \ref{tab: performance}; these speedups are inclusive of the model evaluation during runtime. In the 0-500 MB range, results on Setonix show a higher average speedup performance of 1.32$\times$ compared to 1.07$\times$ on Gadi. In the 0-100 MB memory footprint range, results from Setonix yield an average speedup of 1.41$\times$, and the average speedup on Gadi reaches 1.26$\times$. Generally, we observe a higher standard deviation of speedup in the 0-500 MB range compared to the 0-100 MB range on both Setonix and Gadi; we also observe a higher standard deviation of speedup on Gadi for both ranges. In addition to that, we observe that each percentile's speedup in 0-100 MB is higher compared to 0-500 MB on both Setonix and Gadi, although the 0-500 MB has a higher maximum speedup on both platforms. 

{
For completeness, we report in Table \ref{tab: performance HT off} also the statistics of the ADSALA GEMM speedup results with hyper-threading set to off. The speedup results are largely similar to those using hyper-threading. With hyper-threading set to off, the mean and median speedups on both platforms are slightly lower in the 0-500 MB range, while larger speedups are obtained in 0-100 MB range. The standard deviations of speedups are generally bigger by one to three times except for Gadi in 0-500 MB range, which indicates a more scattered distribution of speedups when hyper-threading is set to off.}


{Following we provide a more detailed analysis and discussion concerning the ADSALA GEMM speedup results using hyper-threading.}

\begin{figure}[!t]
  \centering 
  \vspace{-1em}
     \begin{subfigure}[b]{1\columnwidth}
         \centering
        \makebox[\columnwidth][c]{\includegraphics[width=1.06\columnwidth]{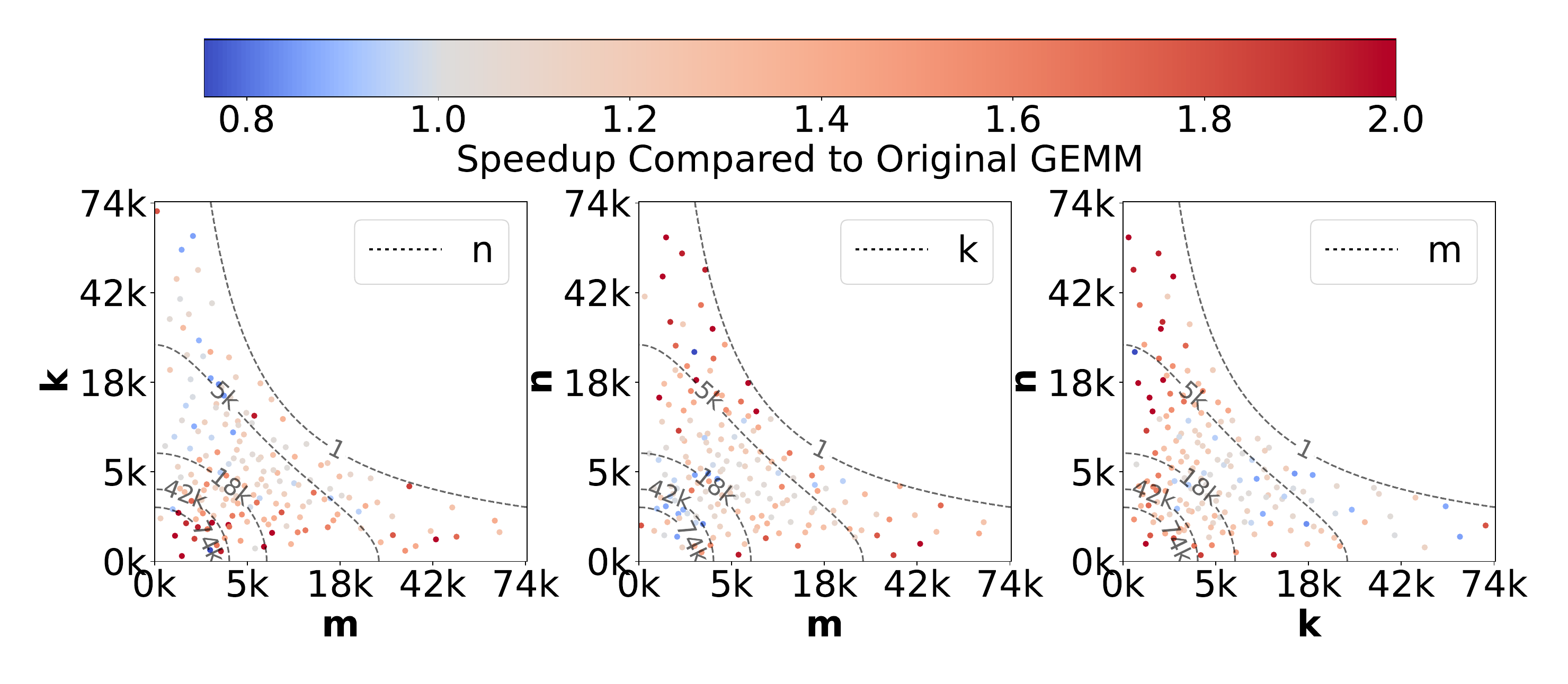}}
        \vspace{-2.2em}
        \subcaption{Setonix}
        \label{fig: Setonix_Speedup_500M} 
     \end{subfigure}
    \vspace{-1.5em}
     \begin{subfigure}[b]{1\columnwidth}
         \centering
        \makebox[\columnwidth][c]{\includegraphics[width=1.06 \columnwidth]{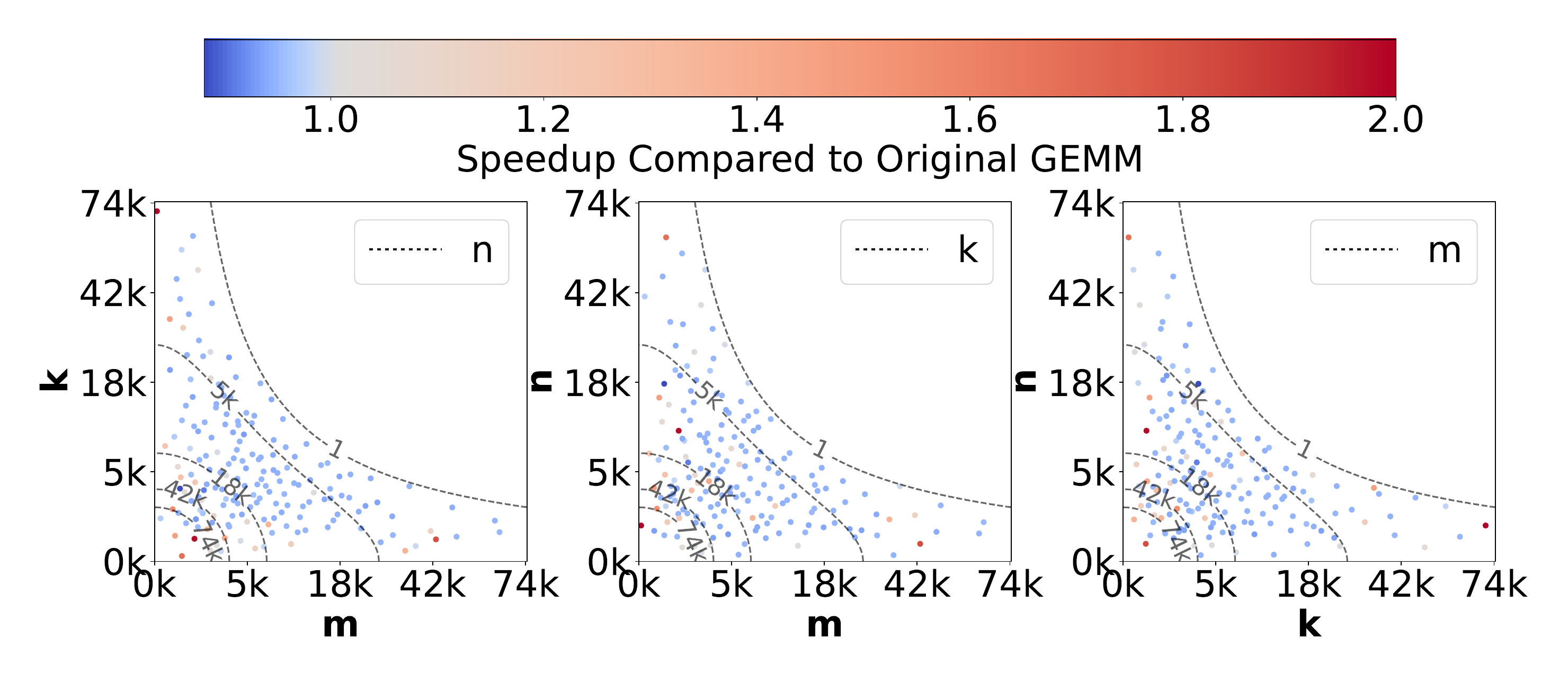}}
        \vspace{-2.2em}
        \subcaption{Gadi}
        \label{fig: Gadi_Speedup_500M} 
        \vspace{1.5em}
     \end{subfigure}

  \caption{Heatmap for speedups with respect to matrix dimensions on Setonix and Gadi. The horizontal and vertical axes use a square root scale. The dashed lines on each sub-graph are contour lines of the sampling domain with each label showing the value of the third dimension.} 
  \label{fig: data collection Gadi} 
   \vspace{-1.0em}
\end{figure}

The distribution of the speedup on Setonix with respect to the matrix dimensions is plotted as a 3D heatmap in Fig. \ref{fig: Setonix_Speedup_500M}. The accelerated GEMM are shown in red, and the decelerated GEMM are shown in blue. It can be observed from the plot that GEMMs with large $n$ values are significantly accelerated, and GEMMs with large $k$ values are moderately accelerated or decelerated. These asymmetric patterns do not perfectly match the patterns observed in the collected data shown in Fig. \ref{fig: data collection Setonix}. Likely, the error from the ML model causes this pattern discrepancy. 

Figure \ref{fig: Gadi_Speedup_500M} shows the speedup visualisation on Gadi. The average speedup is lower than that on Setonix but still significant for GEMM with smaller memory usage, consistent with the observations shown in Fig. \ref{fig: data collection Gadi}. 

\begin{figure}[!tbp]   
\vspace{-2em}  
  \centering
  {\includegraphics[width=0.7\columnwidth]{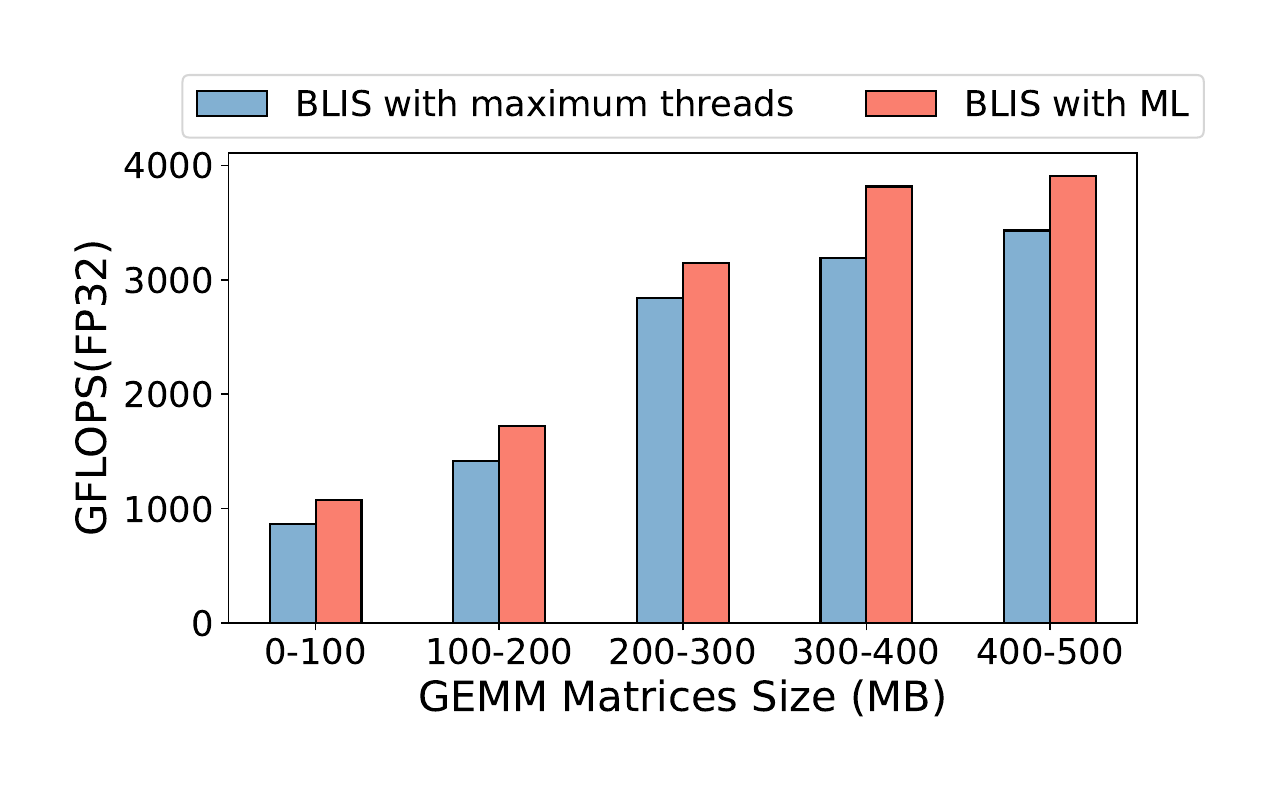}}
    \vspace{-1.5em}
  \caption{GFLOPS performance comparisons with respect to GEMM memory occupancy on Setonix.} 
  \label{fig: Setonix_GFLOPS_500M} 
      \vspace{-1em}
  {\includegraphics[width=0.7\columnwidth]{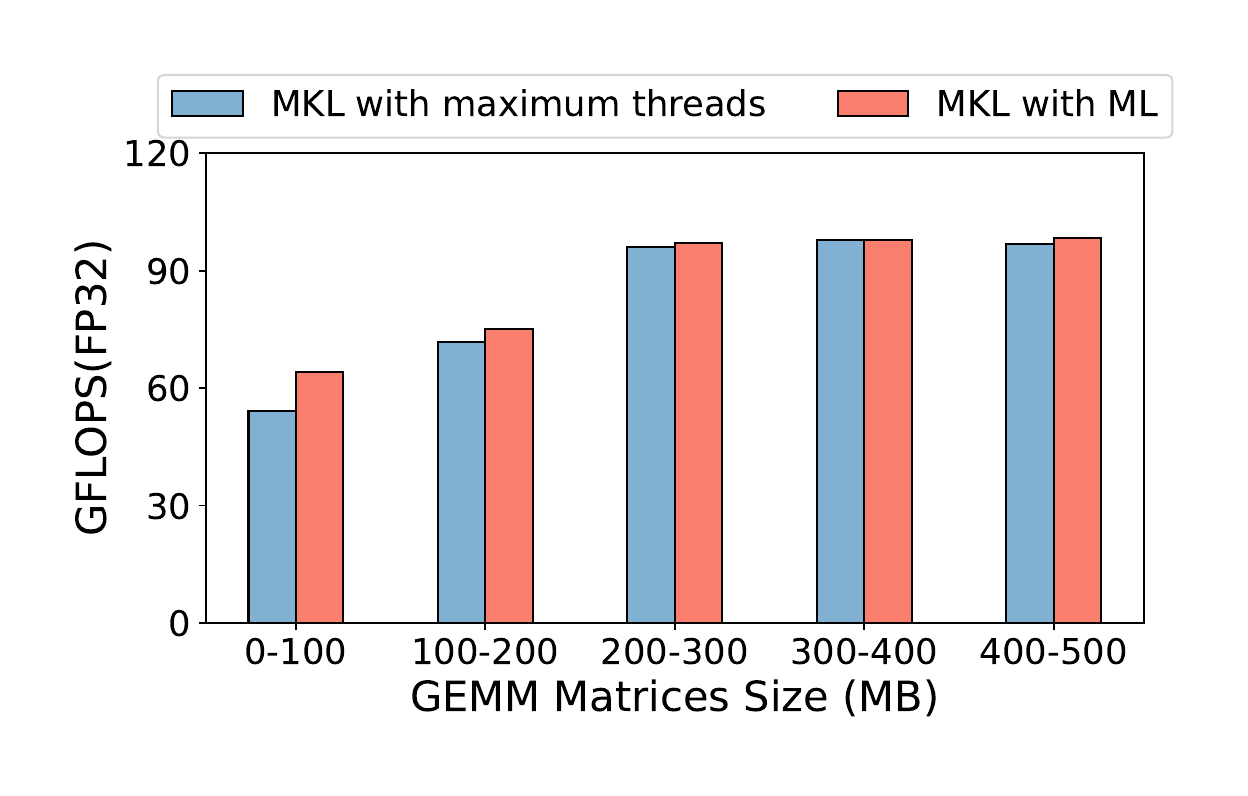}}
    \vspace{-1.5em}
  \caption{GFLOPS performance comparisons with respect to GEMM memory occupancy on Gadi.} 
  \label{fig: Gadi_GFLOPS_500M} 
  \vspace{-1.5em}  
\end{figure}

To better visualise the distribution of the speedup with respect to the GEMM memory occupancy, the performance in GFLOPS is shown in Figures \ref{fig: Setonix_GFLOPS_500M} and \ref{fig: Gadi_GFLOPS_500M}. For GEMMs with memory requirements in the 0-100 MB range, Setonix and Gadi present a similar speedup (around 30\%). 

As the memory requirements increase, the ML-driven thread selection on Setonix shows a relatively stable performance speedup, while on Gadi the speedup tends to converge to one.

\begin{figure}[!tbp] 
\vspace{-2.5em}  
  \centering
  \makebox[\columnwidth][c]{\includegraphics[width=1.05\columnwidth]{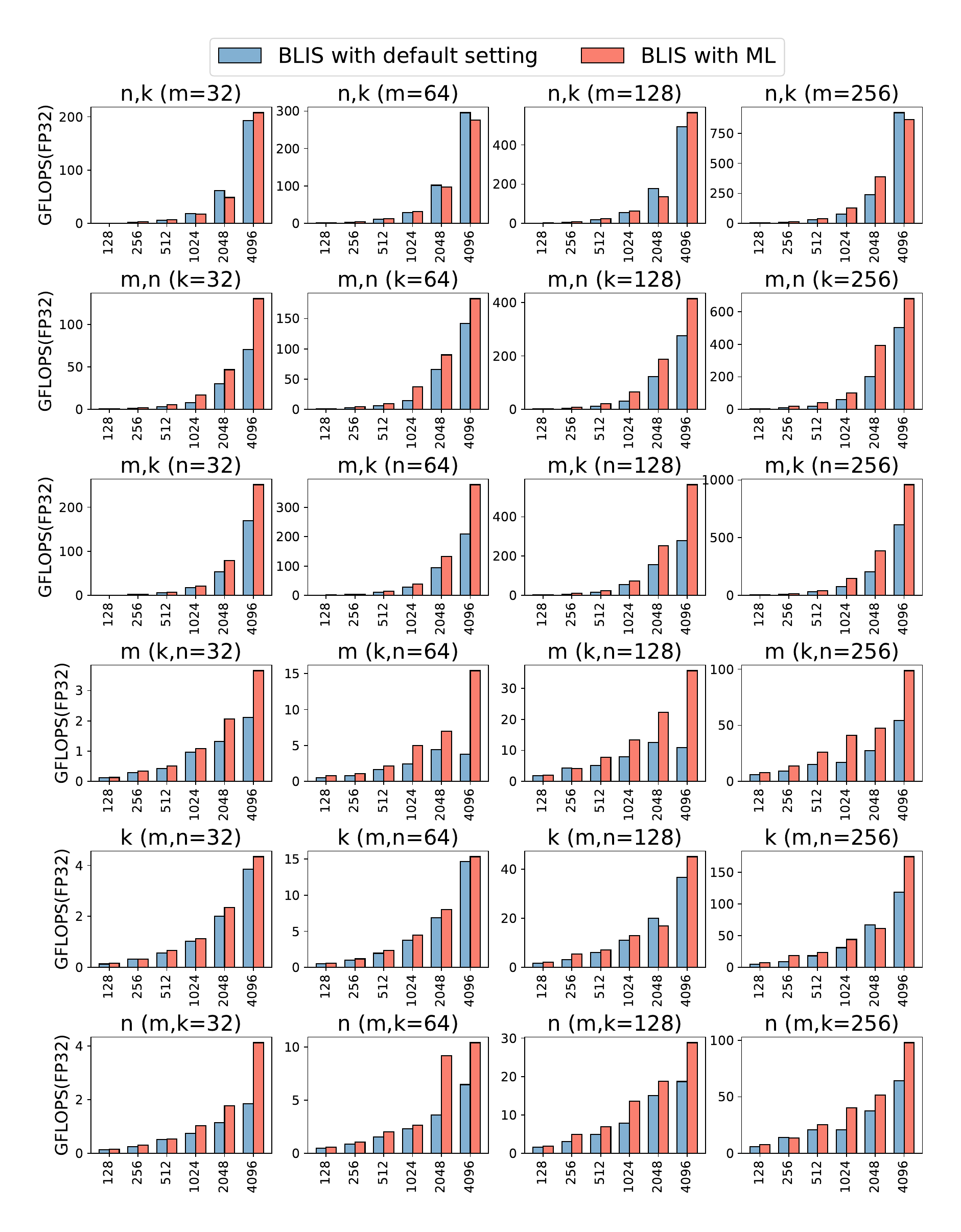}}
  \vspace{-2.5em}  
  \caption{GFLOPS performance comparisons on Setonix with predesigned Matrices} 
  \label{fig: Setonix_irregular_500M} 
    \vspace{-0.5em}  
\end{figure}

\begin{figure}[!tbp]
\vspace{-2.5em}  
  \makebox[\columnwidth][c]{\includegraphics[width=1.05\columnwidth]{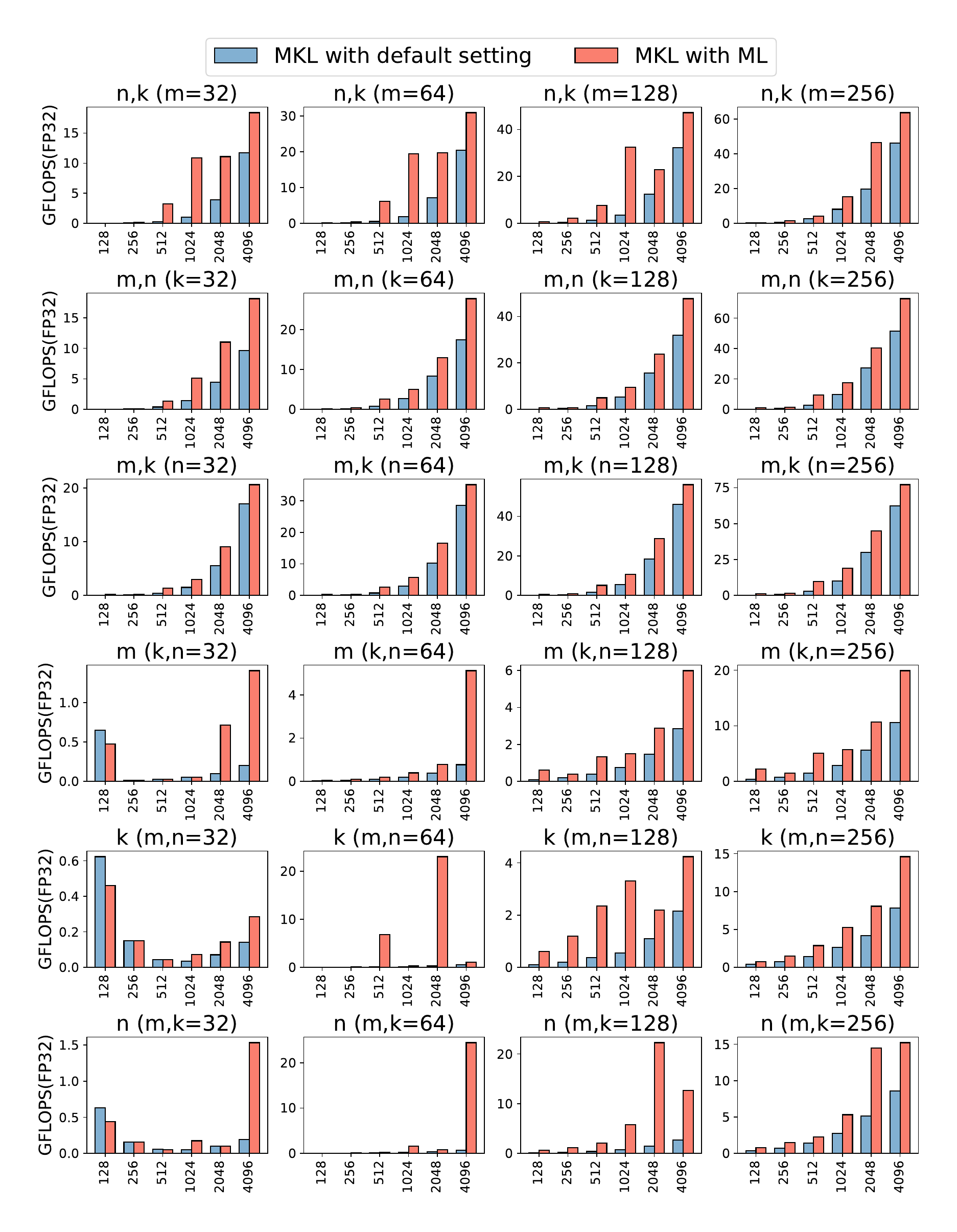}}
\vspace{-2.5em}  
  \caption{GFLOPS performance comparisons on Gadi with predesigned Matrices} 
  \label{fig: Gadi_irregular_500M} 
  \vspace{-1.5em}  
\end{figure}

 Figures \ref{fig: Setonix_irregular_500M} and \ref{fig: Gadi_irregular_500M} show speedup results for ADSALA GEMMs with $m=k=n$, with one small-valued parameter among $m,k,n$ (less than or equal to 256), and  with two small-valued parameters among $m,k,n$. On Setonix, our method demonstrates a stable performance improvement compared to BLIS with the maximum number of available threads. Generally, the speedup increases with respect to the largest dimension(s). As shown in the second and the third rows in Fig. \ref{fig: Setonix_irregular_500M}, when only $k$ or $n$ is small, using ML-based thread selection produces relatively significant speedups, with some of the speedups reaching two. However, as shown in the first row in Fig. \ref{fig: Setonix_irregular_500M}, when only $m$ is small, using ML yields little to slightly adverse speedup in some cases. When $k$, $n$ or $m$, $k$ are small (fourth and the sixth rows in Fig. \ref{fig: Setonix_irregular_500M}), using ML produces relatively high speedups, while when $m$, $n$ are small (fifth row in Fig. \ref{fig: Setonix_irregular_500M}), ML presents a relatively modest speedup.

The test results on Gadi are shown in Fig. \ref{fig: Gadi_irregular_500M}. For some inputs, the original performance is deficient (less than 1 GFLOPS), and our approach gives excellent speedups (\emph{e.g.}, $m=n=64$ and $m=k=64$). Overall, the speedups are unstable across different GEMM inputs, and so is the performance of MKL without ML thread selection. The speedup also does not generally grow with respect to the largest matrix dimensions. When two dimensions are small, corresponding to the last three rows in Fig. \ref{fig: Gadi_irregular_500M}, the performance of MKL is more unstable, with the performance decreasing and then increasing as the input matrix size gets larger. 


Testing on Setonix and Gadi shows that our method can overall and reliably improve the performance of GEMM on different platforms. Also, the complex performance patterns can be captured by the ML model and used by fast runtime evaluations to speed up GEMM workloads significantly.

\subsection{{Profiler-Driven Analysis of Large Speedups}}
{ The results in Fig. \ref{fig: Setonix_irregular_500M} and \ref{fig: Gadi_irregular_500M} occasionally show extraordinary speedups of ADSALA GEMM over traditional GEMM.}

{To explain such large speedups, we used Intel$^{\circledR}$ Advisor and Intel$^{\circledR}$ Vtune on Gadi to perform
detailed profiling of two GEMMs with $m, k, n $ equal to $64,2048,64$ and $64,64,4096$, respectively. As shown in Fig. \ref{fig: Gadi_irregular_500M}, for these two GEMM cases our ML approach yields speedups of $81.6\times$ and $33.9\times$, respectively. To guarantee a sufficient profiler sampling of the GEMM code, each matrix multiplication was repeated 1000 times using different values for the input matrices.}

{Table \ref{tab: profiling Gadi} shows the time breakdown of GEMM calls. According to the profiler results, the parallel SGEMM wall-time has three main components: }

\begin{enumerate}
    \item {Thread synchronization.}
    \item {Data copies. The GEMM runtime maintains a buffer for each thread to use as a workspace; threads copy blocks of matrix operands into their local buffer before operating on them. The amount of copy depends on the sizes of the matrices, their placement in memory, and the number of threads involved. }
    \item {SGEMM kernel calls, wherein all the FLOPs are performed. The specific kernels called depend on the number of threads, as each thread operates on its own blocks of the matrix operands. These can be compute-bound for sufficiently large matrix blocks. }
\end{enumerate}

{For the two cases selected our ML-based SGEMM yields large speedups, consistently with the results shown in Fig. \ref{fig: Gadi_irregular_500M}.  The number of threads selected by the ML model is 14 for the $64,2048,64$ case and 1 for $64,64,4096$. As shown in Table \ref{tab: profiling Gadi}, this results in significant savings across all three components of the wall-time with respect to the 96-thread configuration, with the data copy yielding the largest fraction of time reduction. Note that in the $64,64,4096$ case, when using ML, the thread synchronization and data copy take zero seconds because only one thread is adopted.}

\begin{table}[!htbp]
  \centering
  \vspace{-0.5em}
  \caption{{Profiling results on Gadi with two specific GEMM cases.}}
  \label{tab: profiling Gadi}
  \scriptsize
\sffamily

\begin{tabular}{SSSSSSS}
    \toprule
     {$\textit{\textbf{m, k, n}}$} & \textbf{Thread Count} & \textbf{Total Time (s)} & \textbf{Thread Sync (s)} & \textbf{Kernel Call (s)} & \textbf{Data Copy (s)} \\
    \midrule
    {\textbf{${\text{64,2048,64}}$ no ML}} & 96    &  167.748   &  0.246  &  2.260   & 163.333\\[0.3cm]
    {\textbf{${\text{64,2048,64}}$ with ML}}    & 14    &  1.073     &  0.145  &  0.276   & 0.436\\
    \midrule
    {\textbf{${\text{64,64,4096}}$ no ML}} & 96    & 18.253     &  1.373  &  5.368   & 10.093\\[0.3cm]
    {\textbf{${\text{64,64,4096}}$ with ML}}    & 1     & 0.887      &  0.0     &  0.696   & 0.0\\
    \bottomrule
\end{tabular} 
  \vspace{-1em}
\end{table}

{In summary, these results, which reproduce the large speedups obtained in Fig. \ref{fig: Gadi_irregular_500M}, show that using ML to select a lower number of threads can result in lower data copy and threads synchronization overheads, and in more efficient (higher FLOP rate) GEMM kernel calls, thereby significantly reducing the overall runtime.}




 

\section{Conclusions}
\label{sec: Conclusion}
In this article, we presented a proof-of-concept approach to improving the runtime performance of GEMM routines by using machine learning to automatically select the number of threads that minimizes execution time. 

Results demonstrate suitably selected ML
models can be trained for runtime selection of the optimal number of threads of GEMM and used on-the-fly with minimal overhead to improve performance compared to the default selection of a number of threads equal to the number of physical cores.

More specifically, test results on two different HPC node architectures, one based on a two-socket Intel$^{\circledR}$ Cascade Lake and the other on a two-socket AMD$^{\circledR}$ Zen 3, showed a 25 to 40 per cent average speedup compared to traditional GEMM implementations in BLAS when used as many threads as core available for matrices with an aggregate memory footprint within 100 MB. { Results on other platforms will need to be analysed case by case, but as a general rule  platforms with high CPU core counts can potentially benefit more from ML-based GEMM and for larger aggregate matrix sizes.}

In the future, we plan to extend our ML-driven runtime thread selection approach to other BLAS operations and to a more diverse set of computer systems, including heterogeneous architectures that use a mix of CPUs and accelerators.

\bibliographystyle{IEEEtran}
\bibliography{thesis.bib}

\end{document}